\documentclass[aip,
cha,
preprint,
floatfix,
longbibliography
]{revtex4-2}

\usepackage{amsmath, amssymb}
\usepackage{enumerate}
\usepackage{graphicx}
\usepackage{bm}
\usepackage{hyperref}
\usepackage{nicefrac}
\usepackage{xcolor}
\usepackage[capitalize]{cleveref}
\DeclareUnicodeCharacter{2009}{ }

\usepackage{float}
\makeatletter
\let\newfloat\newfloat@ltx
\makeatother
\usepackage{algorithm, algpseudocode}

\newcommand{\vct}[1]{{\bf #1}}
\newcommand{\nf}[2]{{\nicefrac{#1}{#2}}}

\begin{document}

\title{Introduction to the artificial neural network-based variational Monte Carlo method}
\author{William Freitas}
\affiliation{
Max Planck Institute for the Physics of Complex Systems, 
Noethnitzer Str 38, 01187 Dresden, Germany}

\date{\today}

\begin{abstract}
The construction of trial wave functions based on neural networks combined with the variational Monte Carlo method is discussed. The mathematical formulation for representing quantum states as artificial  neural networks is introduced. The advantages of employing such trial states and how machine learning works are discussed. It is shown that the variational method is a kind of unsupervised learning algorithm, where the multiple minima landscape is used as an asset that leads to a stable optimization procedure. The feature representation plays an important role on interpretability and on extracting physical insights from   nontrivial trial wave functions. The algorithm   is illustrated for the Yukawa potential and the hydrogen molecule.

\end{abstract}

\maketitle

\section{Introduction}

The recent extensive adoption and incorporation of artificial intelligence-based tools in scientific research has led to considerable advancements in several fields. Applications in physics have led to  important results in cosmology,\cite{nta15} particle physics,\cite{gue18} many-body quantum systems,\cite{sar17} quantum chemistry,\cite{pfa20, spe20} statistical mechanics,\cite{zha19} and material science.\cite{wen22} The use of neural networks and other artificial intelligence (AI) tools in physics  is natural because the purpose of these tools is recognizing patterns in data. 

Variational wave functions have  demonstrated exceptional effectiveness in understanding  phenomena such as superconductivity,\cite{bcs57} spin liquids,\cite{and73} and the quantum Hall effect.\cite{lau83} Exploring how neural networks can be used to  build trial wave functions is promising for two reasons. First, neural networks are capable of representing a broad class of functions \cite{cyb89}, including  quantum mechanical wave functions. Variational quantum states that employ neural networks  have been  successful in  investigating  open quantum systems,\cite{nag19} frustrated quantum magnets,\cite{rot23} spin chains,\cite{yan20}  interatomic force estimation, \cite{qia22} excited states of atoms and molecules,\cite{pfa24} homogeneous electron gas,\cite{pes24, cas23} superfluid systems,\cite{lou24} quantum droplets,\cite{fre23, fre24} among others.\cite{car21}

The goal of this article is to provide 
a self-contained  introduction to the artificial neural network-based variational Monte Carlo method (ANNVMC).
The article is organized as follows. Section~\ref{sec:theory} discusses 
the necessary theoretical background to apply ANNVMC, Sec.~\ref{sec:examples} shows the results of applying the method to several quantum systems, Sec.~\ref{sec:remarks} concludes with remarks on this work, and Sec.~\ref{sec:problems} presents some suggested problems. Additional examples \cite{annvmc-github} and a historical background section on the interaction between AI and physics are presented in the supplemental material.\cite{fre26sm}

\section{Theory}\label{sec:theory}

The variational method  relies on the variational principle, which states that for any normalizable function, the expectation value of the Hamiltonian $\mathcal H$ is always equal or larger than the ground state  energy. The method is based on proposing a class of trial wave functions $\psi_\theta(\vct x)$ that depend parametrically on  the parameters $\theta$ and finding the set of optimized parameters that minimizes the expected value of the Hamiltonian. This value is necessarily an upper bound for the true ground state energy and the optimized trial wave function is the best approximation for the ground state within the class of functions $\psi_\theta$. This process requires computing expected value of $\mathcal H$,
\begin{equation}\label{eq:ene}
E[\theta] = \int d\vct x~ p(\vct x) E_L(\vct x)~,
\end{equation}
where
$\vct x = \{q_1, q_2, \ldots, q_{ND}\}$ represents the coordinates of the system, $N$ is the number of particles,  $D$ is the number of spatial dimensions, 
\begin{equation}
p ( \vct x ) = \frac
	{ | \psi_\theta (\vct x)|^2 }
	{ \!\int d\vct x \ | \psi_\theta (\vct x)|^2}  
\end{equation}
is the probability density function,
and  
\begin{equation}
E_L(\vct x) = \frac{\mathcal H~ \psi_\theta(\vct x)}{\psi_\theta (\vct x)} ~
\end{equation}
is the local energy. Multi-dimensional integrals like~\cref{eq:ene} are usually difficult to evaluate. Hence, the Monte Carlo  integration is frequently used to estimate these integrals.

The variational method and  Monte Carlo quadrature are well established methods. A comprehensive description of the variational method is given in Ref.~\onlinecite{cohen} and Monte Carlo quadrature is carefully developed in Ref.~\onlinecite{kalos}.
Variational Monte Carlo employs the variational method  and computes the associated multi-dimensional integrals involved by Monte Carlo quadrature or integration. See Refs.~\onlinecite{tao14,lester} for discussions of this method. The variational Monte Carlo method is also described in the supplemental material.~\cite{fre26sm}

The variational Monte Carlo method is an optimization method that   finds a reasonable approximation for the ground state of a given Hamiltonian. If a proposed trial state has few parameters, one can make a manual search on the parameter space to find the minimum energy. This search quickly becomes impractical as the number of parameters increases, and algorithms with automatic updates for the parameters are needed. A common approach relies on gradient based methods, because the gradient of a function gives the local direction of greatest increase. Therefore, updates in the opposite direction lead to a decreasing value of the energy. The gradient of the energy can be written as
\begin{equation}\label{eq:delE}
\nabla_\theta E[\theta] = 
 \int d\vct x\ p(\vct x)\left(E_L(\vct x) - 
 \int d\vct x' p(\vct x')E_L(\vct x')\right) 
 \nabla_\theta \ln |\psi_\theta(\vct x)|^2 ~.
\end{equation}

Before the raise of neural networks, traditional trial wave functions were proposed based on physical insights and were highly specific and tailored to perform well in a particular system. Usually, proposed functions have specific parametric form that is expected to describe qualitatively the system. On the other hand, neural networks are also parametric functions. However, no assumptions are made about the form of the function that the neural network can represent. Furthermore, the universal approximation theorem~\cite{cyb89} affirms that a sufficiently large neural network can arbitrarily approximate a broad class of functions, which includes normalizable quantum states. Therefore, the neural network-based variational Monte Carlo method proposes that the trial wave function to be a neural network. Due to this flexibility in representing functions, a sufficiently large neural network is able to describe the ground state wave function with high accuracy, while traditional trial functions would be only approximate solutions.

\begin{figure}[tb]
    \centering
    \includegraphics[width=0.85\linewidth]{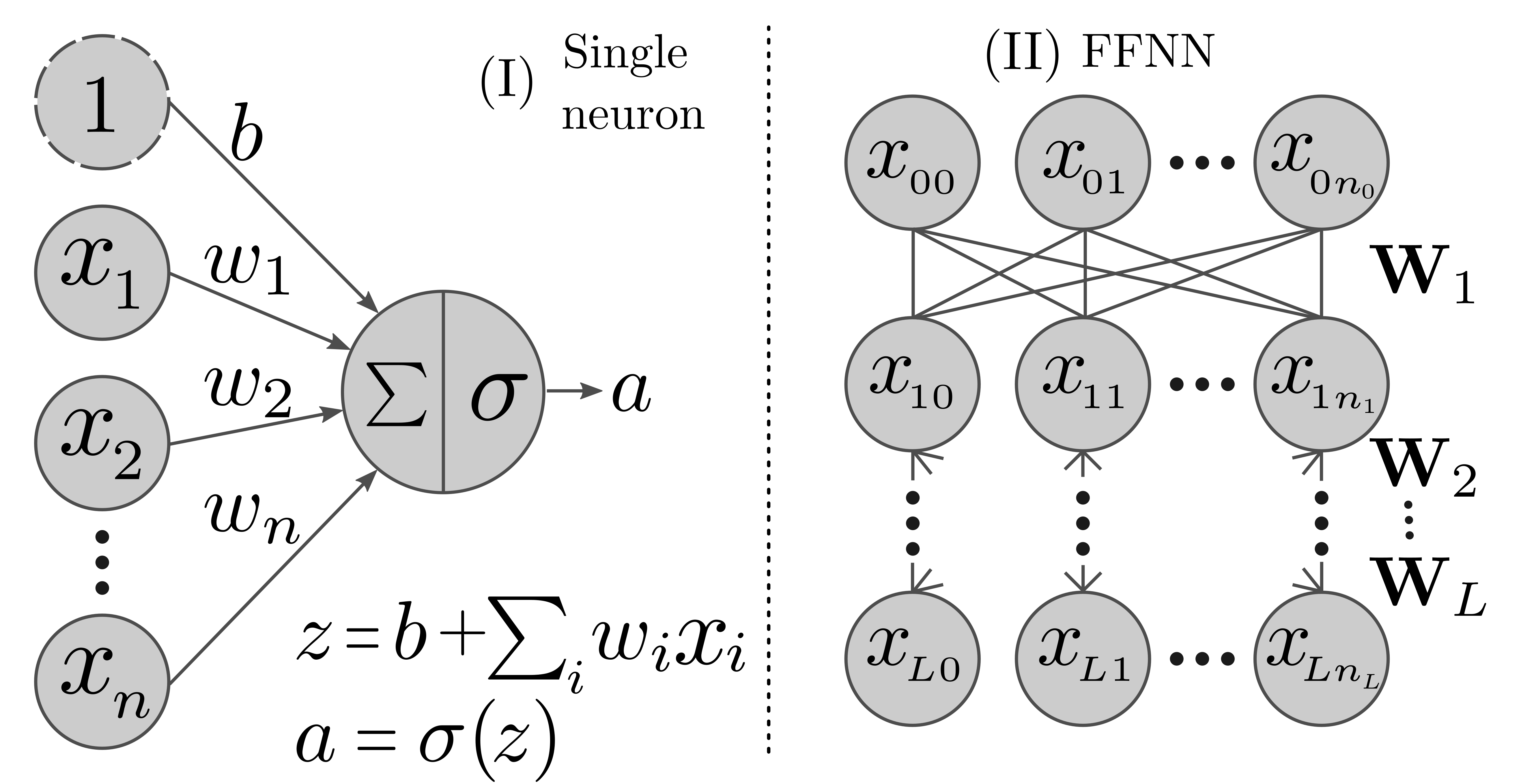}
    \caption{Schematic representations of (I) an artificial neuron  and (II) a feed-forward neural network (FFNN).}
    \label{fig:neuron}
\end{figure}

\textit{Building the trial wave function -} There are many ways to construct neural networks for machine learning algorithms and each  class of tasks usually requires a different architecture. 
Most of them are  based on the same basic units, which are called neurons. Mathematically, these units perform a linear transformation followed by a nonlinear transformation of the inputs. Suppose the inputs are given by the vector $\vct x$, the linear transformation is given by $z=\vct w^\top \cdot \vct x + b$, where the weights $\vct w$ is a vector of parameters   and the parameter $b$ is    called the bias. The nonlinear transformation is given by
\begin{equation}
a = \sigma(z),
\end{equation}
where the nonlinear function $\sigma$ is    the activation function. We will consider the activation function 
\begin{equation}
\sigma(z) = \tanh(z)
\end{equation}
unless otherwise stated. A schematic representation of a neuron is illustrated in Fig.~\ref{fig:neuron}(I).

In order to compose a neural network, the neurons are grouped in layers. They take an input vector and perform the same two transformations, producing an output vector. In other words, if the layer has $m$ neurons and the input vector has dimension $n$, then, the first layer is described by a matrix of weights $\vct W_1$ with $m$ rows and $n$ columns, and also by a vector of biases $\vct b_1$ of size $m$. The index in $\vct W_1$ and $\vct b_1$ indicates the layer number the weights and biases are describing. Thus, the linear transformation is performed by computing $\vct z_1=\vct W_1 \cdot \vct x + \vct b_1$, where $\cdot$ indicates  matrix multiplication. The output vector is denoted by $\vct x_1 = \sigma(\vct z_1)$, where the activation function is applied element-wise to each component of $\vct z_1$. 

Usually, a few layers are stacked one after the other, as it is represented in Fig.~\ref{fig:neuron}(II). To perform this composition, the following recurrence relation is applied 
\begin{subequations}\label{eq:iter-ann}
\begin{align}
\vct z_\ell &= \vct W_\ell \cdot \vct x_{\ell-1} + \vct b_\ell~, \\
\vct x_\ell &= \sigma(\vct z_\ell) ~,
\end{align}
\end{subequations}
where $\vct W_\ell$ and $\vct b_\ell$ are the weights and biases of the $\ell$th layer. The vector $\vct x_0$ represent the input features, in other words, the particle coordinates $\vct x_0\equiv\vct x$. Additionally, in Fig.~\ref{fig:neuron}(II), $x_{\ell i}$ represent the $i$-th component of the vector $\vct x_\ell$. The index $\ell$ are integers in $\{1,\ldots,L\}$ and denotes the layer of the   network, while $\theta=\{\vct W_1,\ldots,\vct W_L,\vct b_1,\ldots,\vct b_L\}$ represents  all parameters of the network. Layers from $\ell=1$ up to $\ell=L-1$ are called hidden layers, and $\ell=L$ is called the output layer. The width of a given layer $\ell$ is defined by the number of rows of the matrix of weights $\vct W_\ell$. The vector $\vct n=\{n_0, ...,n_L\}$ encodes the widths $n_\ell$ for each layer, where $n_0$ is restricted to be the size of $\vct x$. Lastly, $\vct x_L$ are called the neural network outputs.

Note that the output layer is a function of the input features. Namely, for a given configuration in coordinate space $\vct x$ and a set of network parameters $\theta$, the outputs $\vct x_L$ are fully determined by applying Eq.~\eqref{eq:iter-ann}.
Consequently, writing $\vct z_L=\phi_\theta(\vct x)$ explicitly as a function of $\vct x$, the outputs can be written as a composed function $\vct x_L=\sigma(\phi_\theta(\vct x))$. This simple neural network architecture is called feed-forward neural network.
However, the output must be a scalar function to be considered a valid trial wave function. Hence, aiming to fulfil this condition, consider the following neural network-based trial wave function

\begin{equation}\label{eq:psi}
\psi_\theta(\vct x) = \frac 1 {1+e^{\phi_\theta(\vct x)}} ~,
\end{equation}

\noindent
where the constrain $n_L=1$ must be satisfied. Due to how the outputs are computed, $\vct z_L$ must be a vector of size $n_L=1$, which grants that the output is a scalar number. Then, the trial wave function can be computed by applying Eq.~\eqref{eq:psi}. Importantly, the weights and biases condensed in $\theta$ are the variational parameters for the proposed trial function.

Commonly, variational Monte Carlo programs always work with $\ln|\psi_\theta|$ intending to avoid losing numerical precision. Therefore, Eq.~\eqref{eq:psi} employs a
positive function as the activation function for the last layer.
For that reason, it is not required to track the sign of $\psi_\theta$, which facilitates the code implementation. This condition poses no significant limitation for the applications done in this work. Nevertheless, in many systems composed by fermions, the trial wave function must contain negative regions.

\textit{On how to choose the neural network size -} The total number of layers $L$ and the widths $\vct n$ of each layer are user defined quantities. As already mentioned, the two exceptions are $n_0=ND$ and $n_L=1$. As a result, $L$ and $\{n_1,...,n_{L-1}\}$ can be chosen arbitrarily. Any set of widths $\vct n$, that the corresponding neural network is able to properly describe the ground state, is a valid choice. A common option consists of using all hidden layers with the same size, although layers of different widths are also possible.
Nonetheless, if the total number of connections between neurons is smaller than a threshold, the  neural network might lack the necessary flexibility to properly describe the system. Systematically increasing the number of neurons usually leads to better solutions. However the gains of doubling the size of a neural network  become smaller as   the number of neurons increases. If we assume that each hidden layer ($1 \le \ell< L$) to have the same number of neurons, we can increase the size of the network by either fixing the number of neurons per layer or increasing the number of layers. 

Once $L$ and $\vct n$ are chosen, the variational parameters $\theta$ must be initialized carefully to grant stability of the optimization process. Initializing the weights and biases consists in drawing random numbers from a Gaussian distribution of mean zero and unit variance. The weights of a given layer $\ell$ are also divided by the square root of $n_{\ell-1}$. This choice is called  Xavier initialization, which ensures stable signal propagation across layers.\cite{ben10} This procedure is described in Alg.~\ref{alg:initwb}, where ${\rm RandN}(m,n)$ is a function that returns a $m$-by-$n$ matrix of random numbers drawn element-wise from Gaussian distribution and ${\rm RandN}(m)$ does the same for a vector of size $m$.

\begin{algorithm}[ht]
\caption{Function for the initialization of the weights and biases. }\label{alg:initwb}
\begin{algorithmic}[1]
\Require Layer widths $\{n_\ell\}_{\ell=0}^L$
\State $\vct W_\ell \gets {\rm RandN}(n_\ell, n_{\ell-1}) / \sqrt{n_{\ell-1}}$ for $\ell=1,\dots,L$
\State $\vct b_\ell \gets {\rm RandN}(n_\ell)$ for $\ell=1,\dots,L$
\State $\theta \gets  \{\vct W_\ell, \vct b_\ell\}_{\ell=1}^L$
\Return $\theta$
\end{algorithmic}
\end{algorithm}

\textit{Applying variational Monte Carlo with a neural network-based trial function -} As usually done in standard variational Monte Carlo, initial particle coordinates are drawn from random distributions or set according to a rule. In particular, consider that $M$ configurations are initialized and they are denoted by $\{\vct x^{(1)},...,\vct x^{(M)}\}$. At that point, Eq.~\eqref{eq:psi} needs to be evaluated for each configuration $M$ to compute the logarithm of the trial wave function. That is done by applying Eq.~\eqref{eq:iter-ann} iteratively by  standard matrix multiplication, followed by element-wise vector summation, and computing the activation function  $\sigma$ element-wise. For the last layer, instead of applying the activation function, the function to be evaluated is given in Eq.~\eqref{eq:psi}. Then, computing the logarithm of $\psi_\theta$, the result reads $-\ln(1+e^{\phi_\theta(\vct x)})$. This procedure is described in Alg.~\ref{alg:ann}. 

\begin{algorithm}[ht]
\caption{Function describing the computation of $\ln \psi_\theta(\vct x)$.}\label{alg:ann}
\begin{algorithmic}[1]
\Require Input vector $\vct x$ and variational parameters $\theta$
\State $\vct x_0 \gets \vct x$
\For{$\ell = 1$ to $L-1$}
    \State $\vct z_\ell \gets \vct W_\ell \times \vct x_{\ell-1} + \vct b_\ell$
    \State $\vct x_\ell \gets \tanh\!\left(\vct z_\ell\right)$
\EndFor
\State $ z_L \gets \vct W_L \times \vct x_{L-1} + b_L$
\State $y \gets -\ln(1 + e^{z_L})$
\Return $y$
\end{algorithmic}
\end{algorithm}

A set of new configurations are sampled from $p(\vct x)$ by applying the Metropolis algorithm~\cite{met53} every step. After a reasonable amount of steps, the equilibrium distribution of samples is reached. That means new samples are distributed according to $p(\vct x)$. Then, the energy is estimated by

\begin{equation}\label{eq:mc-int}
E[\theta] = \int p(\vct x) E_L(\vct x)~d\vct x \simeq \frac 1 M \sum_i E_L\left(\vct x^{(i)}\right) ~,
\end{equation}

\noindent
which summarizes the Monte Carlo integration. The energy gradients with respect to $\theta$ are  estimated employing the same logic for Eq.~\eqref{eq:delE}. One optimization iteration constitutes of sampling the $M$ configurations and estimating the energy and its gradients. Finally, after estimating $E[\theta_t]$ and $\nabla_\theta E[\theta_t]$ for a given set of $M$ samples in the $t$-th optimization iteration, parameter updates are proposed based on the gradient. The simplest way of updating the parameters using gradient descent is
\begin{equation}\label{eq:update}
\theta_{t+1} = \theta_t - \eta\nabla_\theta E[\theta_t] ~,
\end{equation}
where $\eta$ is the learning rate, a small real number that is  empirically chosen to balance  fast convergence with high stability.  An excessively small value of $\eta$ leads to a very stable optimization procedure, but  very slow convergence. In contrast, a very large $\eta$ might result in noisy energy values and  converge fast initially, but it is likely to drive instabilities or even divergent gradients. The goal is to obtain updates larger than the stochastic noise of the gradient, but smaller than the energy curvature. Values of $\eta$ around $10^{-2}$ or $10^{-3}$ are usual starting points of common optimizers. More advanced details for the optimizer employed in this work are provided in the supplemental material.\cite{fre26sm} Utilizing this optimization procedure, it is possible to find the values of the parameters $\theta$ that minimizes the energy of Eq.~\eqref{eq:ene}. For sufficiently large neural network, this energy minimum is exactly the ground state energy. Consequently, the optimal parameters will describe the ground state wave function through Eq.~\eqref{eq:psi}.

\textit{Local minima in the energy landscape -} Due to non linear activation functions and how the neural network is built, the energy landscape can have several local minima. Since the aim is to find the global minimum, this property could be a problem. However, the energy surface of large neural networks can be approximated by a high-dimensional random energy landscape. Hence, the vast majority of $\nabla_\theta E[\theta]=0$ points are saddle points rather than local minima. The distribution of local minima is highly concentrated near the global minimum so that most minima are nearly equivalent in energy. Also, the number of local minima diminishes exponentially with the size of the neural network. Therefore, the high number of $\theta$ parameters effectively smooths the landscape and creates connected valleys instead of isolated local minima, which leads to many parameter configurations representing approximately the same function. That leads to the existence of regions dominated by plateaus in the energy that are easier to reach and more stable to noise. As a consequence, these characteristics facilitate efficient optimization and prevent the process from becoming trapped in suboptimal local minima. Furthermore, because the minima are connected by low barriers instead of isolated wells, the relatively rare cases where the energy is in a suboptimal local minimum, the stochastic nature of the procedure means that the gradients are noisy, which helps the energy to escape such regions.

\textit{Comparison with machine learning -} In contrast to what was described for the variational Monte Carlo, many learning algorithms require labeled data, which act as guide  to aid the  learning of neural networks. In other words, the user must provide a known data set with expected correct answers. Then, the algorithm tries to make predictions and compare them with the correct answers. The network parameters are updated based on its correctness and, therefore, the learning is supervised. On the other hand, variational methods do not use any labeled data. Rather, the trial function is interpreted as a probability density from which the particle coordinates are drawn from. Based on the energy, that depends on the sampled coordinates, the network adjusts itself to improve the energy. In this scenario, the energy is a feedback and, because there is no labeled data to compare with, this class of algorithms is called unsupervised learning. Note that unsupervised learning is a broader class of algorithms, but for the purposes of this paper, our definition is adequate.

The neural network based variational Monte Carlo method can be interpreted as an exact solver for the Schr\"odinger equation that also produces an analytical form for the ground state wave function, in contrast to other numerical solvers such as finite-difference methods that usually requires interpolation. Choosing the input features to be  particle coordinates is the most basic and fundamental information needed to describe a wave function. Nevertheless, it is one among many possibilities of feeding the neural networks with information about the system. For a single particle in a central potential, for example,      the distance $r=\sqrt{x^2+y^2+z^2}$ is important and  can also be used as information to feed the neural network. By systematically varying the set of input features, it is possible to assess how the trial function performance depends on the  representation chosen. Such comparisons can provide physical insight by revealing which features capture the essential degrees of freedom of the system. Hence, the  performance acts as a probe of the physically meaningful degrees of freedom. Also, because the optimized neural network describes the ground state, additional insights can be extracted from its analytical form.

\section{Examples}\label{sec:examples}

Applying the variational Monte Carlo method requires to define the number of samples $M$ used in Eq.~\eqref{eq:mc-int}  and how many optimization iterations $N_{\rm OI}$ to perform. The value of $M$ defines how many samples are used to estimate the  energy. Hence, it should be large enough to provide meaningful values. Usually, $M$ values are on the order of $10^3-10^5$. In contrast, $N_{\rm OI}$ can be  increased until the simulation  converges. For this work, each simulation considered $M=4096$ samples per iteration and, a total of $N_{\rm OI} = 5000$ optimization iterations. 

In addition to the pseudo algorithms described on Sec.~\ref{sec:theory}, a standard variational Monte Carlo  algorithm is also included in Ref.~\onlinecite{fre26sm}. The Introduction to the artificial neural network-based variational Monte Carlo program~\cite{annvmc-github} employed for the simulations in this work is also publicly available and can be consulted as a reference implementation. 

\subsection{Yukawa potential}

The Yukawa potential is
\begin{equation}
V(\vct r) = -\alpha ~\frac{e^{-\nf r \gamma_d}}r~,
\end{equation}
where $\gamma_d$ is a potential parameter and $r=|\vct r|$ is the particle distance. For a particle of mass $m$ and length units given by the Bohr radius $a_{\rm B}=\hbar/mc\alpha$, where $\alpha$ is the fine-structure constant and $c$ is the speed of light, the energy  is expressed in terms of the  Hartree energy $\hbar^2/m a_{\rm B}^2$. If we expand the analytical solution to order $\delta^3$, where $\delta=a_{\rm B}/\gamma_d$, the ground state wave function is $\psi(\hat r) \propto e^{-r+\frac 1 {12}\delta^2(3-2\delta)r^2 - \frac {\delta^3}{18} r^3}$, and the dimensionless ground state energy is $- 1/2 (1-2\delta+3\delta^2/2 -\delta^3)$. This expansion is reasonable for $\delta \lesssim 0.2$ or smaller values~\cite{rod21} and is the value considered in the simulation.

The input features are the spatial coordinates of the electron, namely $\vct x\equiv\vct r$. Initially, the $M$ samples are drawn from a radially uniform distribution of unit radius  centered at the origin. Other kinds of initializations for the coordinates are also possible, such as using a uniform distribution in a box. After choosing the number of layers $L$ and their widths $\{n_1,...,n_L\}$, the weights and biases are initialized according to Alg.~\ref{alg:initwb}. In sequence, the log probability $\ln|\psi(\vct x)|$ is computed using Eq.~\eqref{eq:psi} as described in Alg.~\ref{alg:ann} for each sample. Then, a trial move is proposed for each sample and accepted or rejected as done in the Metropolis algorithm. With the new samples, the energy and its gradients are estimated employing Monte Carlo integration for Eq.\eqref{eq:ene} and Eq.~\eqref{eq:delE}, respectively. Lastly, the parameters are updated according to Eq.~\eqref{eq:update}. After that, the process is repeated for $N_{\rm OI}$ iterations. Those steps summarize the standard variational Monte Carlo method employing a trial wave function that is a neural network. The acceptation rate of proposed trial moves is determined by the size of the trial step. In the beginning of the simulation,  the distribution of samples are not equilibrated, and a small trial step favors the moves to be accepted more frequently and, therefore, helps the sampling to equilibrate. Hence, the trial step is adjusted adaptively according to the acceptance rate, where at the beginning of the simulation, a rate close to 99\% is appropriate, whereas at later stages a value of approximately 50\% is typically recommended.

\begin{figure}[t]
    \centering
    \includegraphics[width=0.85\linewidth]{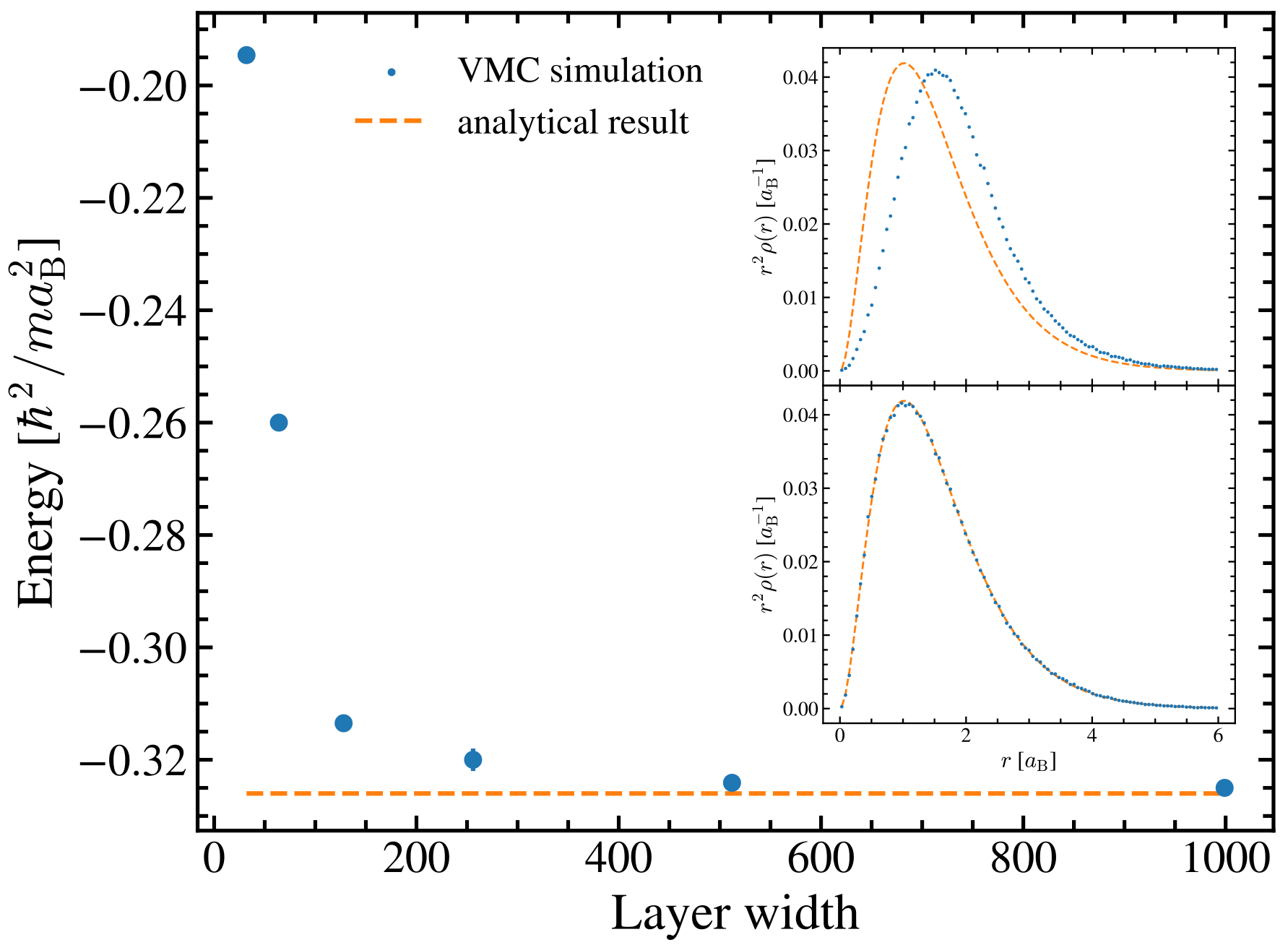}
    \caption{Blue dots showing the optimized energy values as a function of the layer width for a single hidden layer  neural network. The insets show the estimated probability density function  for $k=32$ (top) and $k=512$ (bottom) neurons. Analytical results are displayed in orange dashed lines for reference. }
    \label{fig:ene-vs-width}
\end{figure}

The number of layers $L$ and their widths $n_\ell$ that define the  neural network size are  flexible  for the reasons we have  mentioned. Therefore, analyzing the behavior of the energy as a function of  size is crucial to identify the ground state energy properly. Consider a $L=2$  with $n_1=k$, where the $k$ values considered are 32, 64 128, 256, 512, and 999. For each value of $k$, a variational Monte Carlo optimization was conducted and the final energy results are displayed in Fig.~\ref{fig:ene-vs-width}. As the    size increases, the energy approaches the analytical dimensionless energy of $-0.326$. For smaller $k$ values, the simulation could not reach bound states, which was evident because all kinetic, potential, and total energies were zero.
Additionally, the insets of Fig.~\ref{fig:ene-vs-width} also present a comparison between analytical radial probability density function $\rho(\hat r)=|\psi(\hat r)|^2$ and the stochastic sampled probability density function for $k=32$ (top panel) and $k=512$ (bottom panel). The discrepancy from the analytical solution showed by the neural network on the top panel confirms that the width is too small, while the bottom panel demonstrate that for $k\ge512$, the width is enough to reach near exact solutions. 

\begin{figure}[tb]
    \centering
    \includegraphics[width=0.85\linewidth]{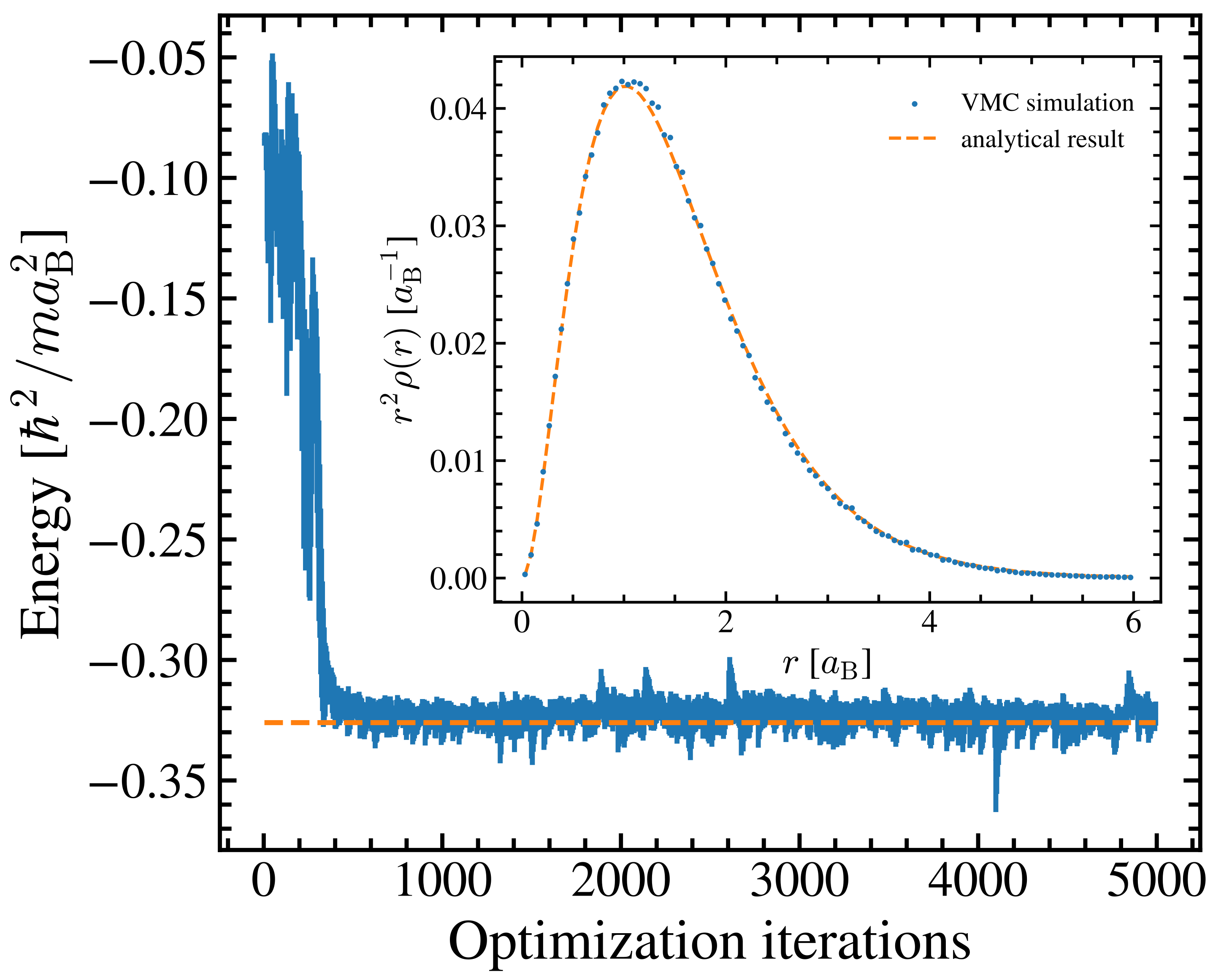}
    \caption{Neural network-based trial state description of the Yukawa potential. The main graph displays the energy as a function of the optimization iterations, and the inset compares the radial probability density function sampled in the simulation with the analytical probability density function up to order $\delta^3$. }
    \label{fig:opt-yu}
\end{figure}

A typical curve showing the evolution of the variational energy as function of the optimization iterations is shown in Fig.~\ref{fig:opt-yu}, where the error bars are also included. For this simulation, the employed neural network architecture is composed by $L=4$ layers, where the layers $[1,2,3,4]$ were formed by $[24,32,16,1]$ neurons, respectively. These specific layer widths were chosen to demonstrate that width values are flexible. Any set of values that is able to reach ground state accuracy is satisfactory. Results obtained in this work are shown in blue dots and analytical results in dashed orange lines. 
The energy quickly converges to the analytical result and oscillates around it. Note that, because there are more layers, a considerable less amount of neurons are needed to reach the ground state. Because the neurons are fully connected between adjacent layers, the total number of connections is the product of the widths. This is effectively similar to a single hidden layer with a much larger width, because one hidden layer produces only one connection per neuron with the output layer. Furthermore, a comparison between the analytical and stochastic probability density function is also presented in the inset, demonstrating an excellent agreement.

\begin{figure}[tb]
    \centering
    \includegraphics[width=0.85\linewidth]{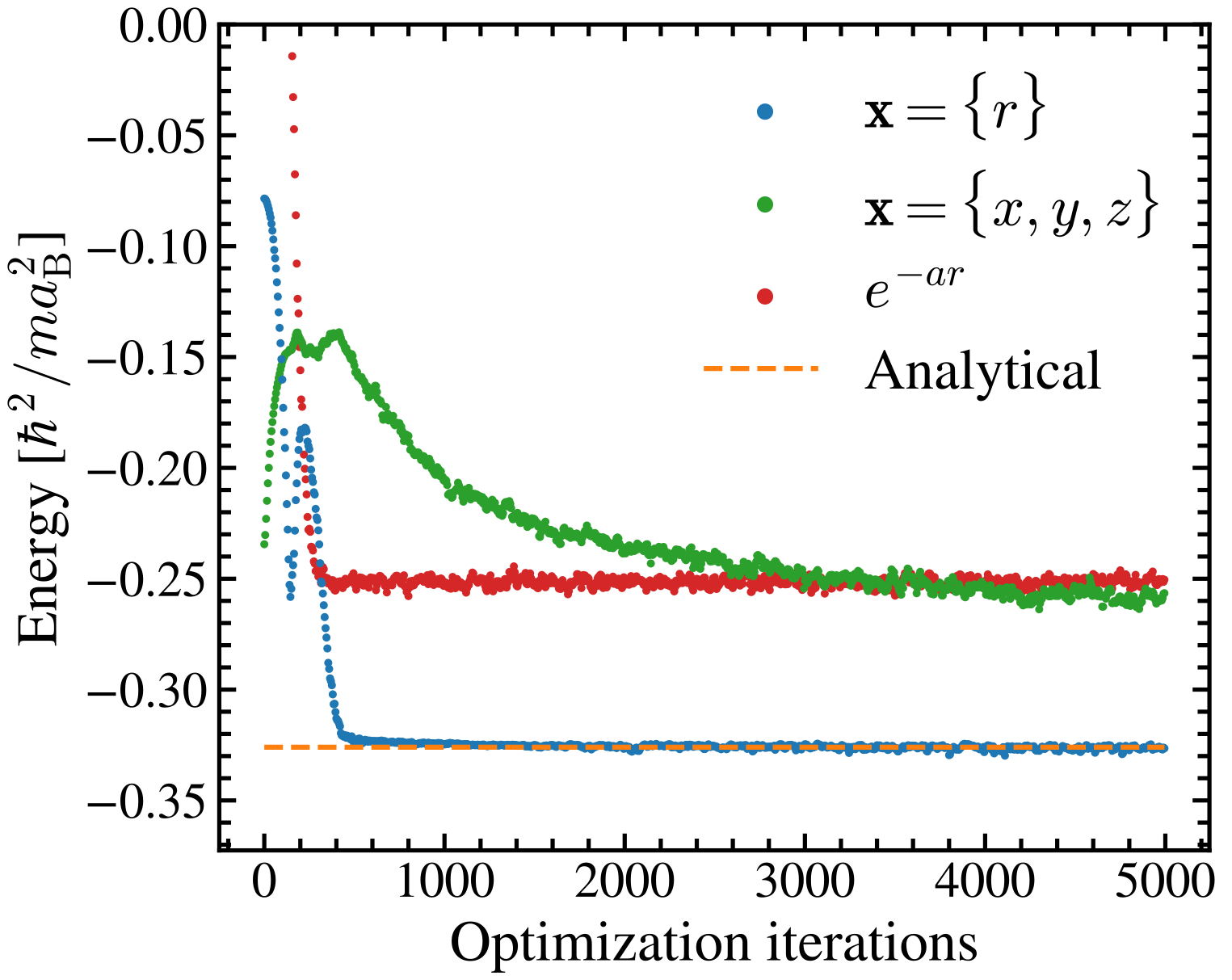}
    \caption{Comparison between the energy as a function of the optimization iterations of neural networks containing hidden layers of equal size but different sets of input features. }
    \label{fig:opt-features}
\end{figure}

Because the Yukawa potential depends only on the particle distance $r$, the system is rotationally invariant. Consequently, the radial and angular degrees of freedom are decoupled. Therefore, to determine the ground state wave function, the distance $r$ is the only  variable needed to describe the state. For that reason, a simulation considering $\vct x=\{r\}$ only  as input feature was performed. A  neural network with a single hidden layer and 64 neurons was employed ($L=2$ and widths given by [64,1]). The result is    compared with the simulation result showed in Fig.~\ref{fig:ene-vs-width}. Figure~\ref{fig:opt-features} displays the evolution of the energy in the optimization process as a function of the iterations for both  neural networks. When using Cartesian coordinates, the number of neurons is still small and the procedure is not able to reach ground state accuracy. In contrast, the reference energy is quickly reached when applying the radial coordinate as input feature, even for the same number of neurons. The oscillations are also less pronounced when considering a better representation for the input features. Hence, we see  the importance of representing well the input features and that the most relevant variable is the radial distance $r$. Additionally, Fig.~\ref{fig:opt-features} also shows the evolution of the variational energy for the hydrogen-like orbital $e^{-ar}$ as trial function, where $a$ is the variational parameter. The difference in performance between a traditional trial function and a neural network-based trial function is evident.

\subsection{Hydrogen molecule}

Lastly, the hydrogen molecule $H_2$ was studied. The protons are considered static in their equilibrium position and separated by $R=1.399~a_0$. The resulting Schrodinger equation describes the behavior of the $N=2$ electrons and, therefore, the total potential of the system is written as
\begin{equation}
V(\vct x) = \frac{e^2}{4\pi\epsilon_0} \left[ - \frac {1}{\left|\vct r_1 - \nf{\vct R}2\right|} -
\frac {1}{\left|\vct r_1 + \nf{\vct R}2\right|} 
- \frac {1}{\left|\vct r_2 - \nf{\vct R}2\right|} -
\frac {1}{\left|\vct r_2 + \nf{\vct R}2\right|} +
\frac {1}{|\vct r_1 - \vct r_2|} +
\frac {1}{R} \right]~,
\end{equation}
where $\vct R=R~\hat z$ is the vector that connects the two protons. Since there are $N=2$ particles in $D=3$ spatial dimensions, $n_0=6$. The input features are defined by the concatenation of particle coordinates $\vct x\equiv\{\vct r_1,\vct r_2\}$. In order to reach ground state level accuracy, the neural network architecture employed was enlarged, making use of $L=4$ layers with $[64, 64, 64, 1]$ neurons for each respective layer. Moreover, the system is formed by two fermions and, hence, it requires the wave function to be antisymmetric. For the ground state, the system is in a singlet spin state, which is antisymmetric. Accordingly, the spatial wave function must be symmetric. Nevertheless, that property is not imposed to the neural network trial state. Instead, the training process is responsible to learn this feature.

\begin{figure}[tb]
    \centering
    \includegraphics[width=0.85\linewidth]{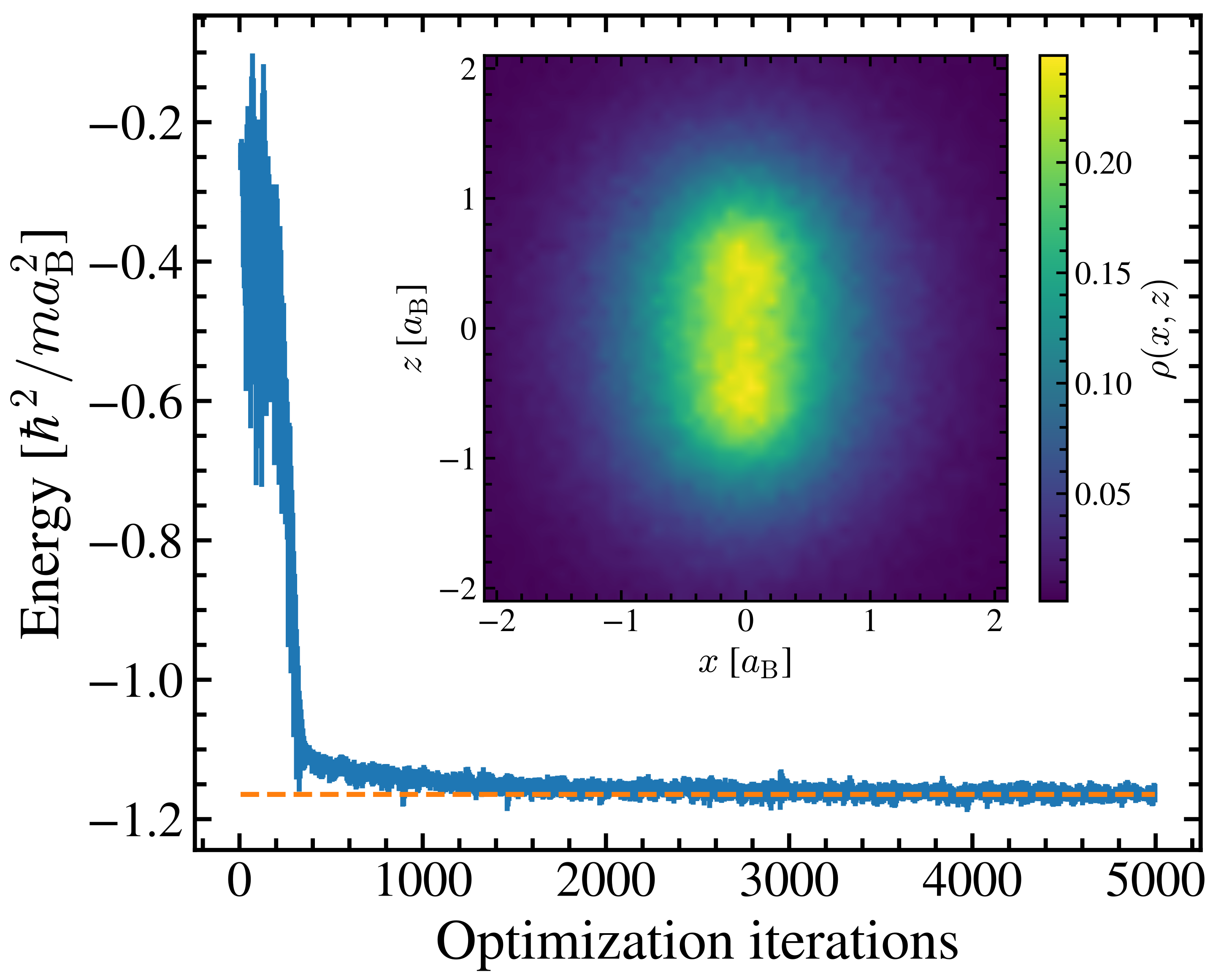}
    \caption{The minimization process for the H$_2$ shows the energy as a function of optimization iterations. The inset present a 2D projected distribution function.}
    \label{fig:opt-h2-mol}
\end{figure}

In this scenario, the numerical available energy result predicts -1.1645 Hartree for the ground state energy~\cite{sch96}. The progression of the variational energy as a function of the number of optimization iterations is displayed in~ Fig.~\ref{fig:opt-h2-mol}. Initially, the procedure shows a quick reduction on the energy estimation followed by a slow decay towards the reference energy. By the end of the procedure, the energy is oscillating around the predicted value. Additionally, the projected probability density function is also shown in the inset. 

The accuracy achieved by the neural network trial states is remarkable, which reinforces the large versatility within a single parametrized function. It is also important to point out that this architecture does not implement any physical feature in the trial state. By naively offering $\{q_1,\ldots,q_{ND}\}$ all component coordinates as an input for the neural network, properties such as cusp conditions and spin symmetry have to be learned by the neural network. Even without these features built in the architecture, the learning process was able to capture the ground state accurately for these simple systems.

\section{Final remarks}\label{sec:remarks}
The neural network based trial states, as proposed in this work, are quite limited. Due to the simplicity and lack of physical insights of the Ansatz, high accurate results are restricted by the number of degrees of freedom $n_0$. Considering that the particles interact, as $n_0$ grows, the neural network size needed to reach ground state accuracy also grows, as well as the need to perform more optimization steps. As previously mentioned, the universal approximation theorem grants that an infinitely large neural network can capture the ground state wave function of any system, regardless of the simplicity of the trial state. However, that is limited due to computer power and simulation time. Therefore, there is a necessity of going beyond the simplest neural network architecture for more complex systems. To achieve that, the input features for the neural network can be improved by employing more relevant quantities than the naive plain particle coordinates $\vct x$. One possible example would be utilizing the relative vector coordinates for multi particle systems interacting through pair potentials. Another possibility would be using the relative distances together with, considering the Born-Oppenheimer approximation, the particle distances to the nuclei in a molecule. Each class of systems have different quantities that are important for the description of the ground state. That also can be used to extract physical information or insights from the systems, because the performance of a given set of input features is directly correlated with how the features encode the relevant quantities of such a system.

Furthermore, depending on the system, symmetries, such as global translation invariance, can even make the training unfeasible if the trial state is built without considering this property.\cite{fre23} Other examples~\cite{hol18,pfa20,car21} of properties that can be added to significantly improve the training of neural networks include spin statistics, cusp conditions, periodicity, backflow and equivariant transformations. Overly distinct system properties requires tweaks in the architecture and input features that are provided to the neural network. However, class of systems with similar properties can be described by the same architecture, even though they can be intrinsically distinct. For example, electronic structure of circular quantum dots~\cite{fre25} and Bicyclobutane molecule~\cite{spe20} use the same architecture because they belong to the same class of systems, that is, finite size spin $\nf 1 2$ fermions. Despite system dependent adjustments being required sometimes, the general structure of stacking layers of artificial neurons is preserved and employed across different applications successfully. 

Finally, neural networks based trial wave functions have been proving to be able to describe quantum states across several systems with distinct features. The flexibility, performance and the generalization character of such trial functions show how powerful is this tool and it demonstrates that they can aid the development of physics research. Nevertheless, there are other approaches and methods that also employ concepts borrowed from the AI field to study physics. For example, one perspective would be to study ultracold molecules by means of Gross–Pitaevskii equation associated with neural networks. 

\section{Suggested problems}\label{sec:problems}

\begin{enumerate}
  \item Derive the relation of Eq.~\eqref{eq:delE} by taking the derivative with respect to the variational parameters $\theta$ of the expected value of the Hamiltonian for a generic real trial wave function $\psi_\theta(\vct x)$. Hint: remember that the normalization constant of the trial wave function depends on its variational parameters.

  \item Find the optimized energy for the Hydrogen molecule ion $H_2^+$ (more details on the system are found in Ref.~\onlinecite{fre26sm}) by employing an  neural network of $L=2$ layers as the trial wave function, where the number of neurons in each layer is given by $[k,1]$. Compare the energy results with the dimensionless value $E_0=-0.59724$.\cite{sch96} Start with a small value of $k$ and increase it until the optimized energy changes less than 0.1\%. After that, do the same for $L=3$ ($[k,k,1]$ neurons for each layer) and $L=4$ ($[k,k,k,1]$). Compare the total number of neurons necessary to reach ground state accuracy for the different number of layers. \label{prob:2}

  \item Find the optimized energy of the Helium atom that is governed by the dimensionless Hamiltonian
  \begin{equation}
  \mathcal H=-\nf {\nabla_{\vct r_1}^2}{2}-\nf {\nabla_{\vct r_2}^2}{2} - \nf 2 {r_1} - \nf 2 {r_2} +\nf 1 {|\vct r_1 - \vct r_2|}
  \end{equation}
by employing a  neural network as the trial wave function. Start by considering a $L=4$ layer  neural network with the number of neurons in each layer given by $[2^k, 2^k, 2^k, 1]$, where $k$ is a   to be determined. For reference, in atomic units, the ground state energy of the Helium atom~\cite{nist-he} is $E_0=-2.9034$. Find the smallest value of $k$ that can achieve ground state accuracy. Propose a scaling power law to describe the  of the optimized energy as a function to the number of neurons.

  \item Simulate the hydrogen molecule again as described in Sec.~\ref{sec:examples}(B), but, instead of using the electron coordinates as input, include only the electrons distance to both nuclei and their relative distance. That is, the input features were $\{x_1,y_1,z_1,x_2,y_2,z_2\}$ in Sec.~\ref{sec:examples}(B), while for this  problem choose the input features as $\{|\vct r_1+\nf {\vct R} 2|, |\vct r_1-\nf {\vct R} 2|, |\vct r_2+\nf {\vct R} 2|, |\vct r_2-\nf {\vct R} 2|, |\vct r_1-\vct r_2|\}$. Compare the results for   networks of approximately the same size, namely, equivalent number of neurons in each layer for both   inputs.

  \item Calculate the ground state energy for a one-dimensional system  of four particles trapped in a harmonic oscillator potential and interacting pairwise through a Gaussian potential. The dimensionless Hamiltonian is given by
  \begin{equation}
  \mathcal H= -\nf 1 2\sum_i \nf {\partial^2}{\partial x_i^2} + \sum_i\nf {x_i^2} 2 + V_0 \sum_{i<j} \exp [-\nf {|x_i-x_j|^2}{2\sigma^2}].
\end{equation}
 Choose  the Gaussian width $\sigma=0.1$. Compare the results for simulations for $V_0=10$ and $V_0=-10$. In particular, investigate the differences in the one-body density $n(x)=\int|\psi(x,x_2,x_3,x_4)|^2~dx_2dx_3dx_4$. This function is proportional to the histogram of individual particles coordinates that are sampled throughout the Monte Carlo integration after the optimization process. Check that the solution reduces to the non interacting harmonic oscillator as $V_0\to 0$.
  
\end{enumerate}

\section*{Acknowledgments}


The author wishes to thank Prof.\ Axx for fruitful discussions and Prof.\ Bxx for the encouragement.

The author has no conflicts to disclose.

\bibliography{ref}

@article{turing50,
    author = {TURING, A. M.},
    title = {I.—COMPUTING MACHINERY AND INTELLIGENCE},
    journal = {Mind},
    volume = {LIX},
    number = {236},
    pages = {433-460},
    year = {1950},
    month = {10},
    issn = {0026-4423},
    doi = {10.1093/mind/LIX.236.433},
    url = {https://doi.org/10.1093/mind/LIX.236.433},
    eprint = {https://academic.oup.com/mind/article-pdf/LIX/236/433/30123314/lix-236-433.pdf},
}

@book{kalos,
  title={Monte carlo methods},
  author={Kalos, Malvin H and Whitlock, Paula A},
  year={2008},
  publisher={John Wiley \& Sons}
}

@misc{nist-he,
author = {A.~Kramida and {Yu.~Ralchenko} and
J.~Reader and {and NIST ASD Team}},
HOWPUBLISHED = {{NIST Atomic Spectra Database
(ver. 5.12), [Online]. Available:
{\tt{https://physics.nist.gov/asd}} [2026, March 29].
National Institute of Standards and Technology,
Gaithersburg, MD.}},
year = {2024},
}

@InProceedings{ben10,
  title = 	 {Understanding the difficulty of training deep feedforward neural networks},
  author = 	 {Glorot, Xavier and Bengio, Yoshua},
  booktitle = 	 {Proceedings of the Thirteenth International Conference on Artificial Intelligence and Statistics},
  pages = 	 {249--256},
  year = 	 {2010},
  editor = 	 {Teh, Yee Whye and Titterington, Mike},
  volume = 	 {9},
  series = 	 {Proceedings of Machine Learning Research},
  address = 	 {Chia Laguna Resort, Sardinia, Italy},
  month = 	 {13--15 May},
  publisher =    {PMLR},
  url = 	 {https://proceedings.mlr.press/v9/glorot10a.html}
}

@article{tao14,
    author = {Pang, Tao},
    title = {Diffusion Monte Carlo: A powerful tool for studying quantum many-body systems},
    journal = {American Journal of Physics},
    volume = {82},
    number = {10},
    pages = {980-988},
    year = {2014},
    month = {10},
    issn = {0002-9505},
    doi = {10.1119/1.4890824},
    url = {https://doi.org/10.1119/1.4890824},
}

@book{lester,
  title={Monte Carlo methods in ab initio quantum chemistry},
  author={Hammond, Brian L and Lester, William A and Reynolds, Peter James},
  volume={1},
  year={1994},
  publisher={World Scientific}
}

@article{hil07,
  title={To recognize shapes, first learn to generate images},
  author={Hinton, Geoffrey E},
  journal={Progress in brain research},
  volume={165},
  pages={535--547},
  year={2007},
  publisher={Elsevier},
  url={https://api.semanticscholar.org/CorpusID:10327263}
}

@article{pes24,
	title = {Message-passing neural quantum states for the homogeneous electron gas},
	volume = {110},
	url = {https://link.aps.org/doi/10.1103/PhysRevB.110.035108},
	doi = {10.1103/PhysRevB.110.035108},
	number = {3},
	urldate = {2025-06-11},
	journal = {Phys. Rev. B},
	author = {Pescia, Gabriel and Nys, Jannes and Kim, Jane and Lovato, Alessandro and Carleo, Giuseppe},
	month = jul,
	year = {2024},
	note = {Publisher: American Physical Society},
	pages = {035108},
}

@article{lou24,
	title = {Neural {Wave} {Functions} for {Superfluids}},
	volume = {14},
	url = {https://link.aps.org/doi/10.1103/PhysRevX.14.021030},
	doi = {10.1103/PhysRevX.14.021030},
	number = {2},
	urldate = {2025-08-07},
	journal = {Phys. Rev. X},
	author = {Lou, Wan Tong and Sutterud, Halvard and Cassella, Gino and Foulkes, W. M. C. and Knolle, Johannes and Pfau, David and Spencer, James S.},
	month = may,
	year = {2024},
	note = {Publisher: American Physical Society},
	pages = {021030},
}

@article{cas23,
	title = {Discovering {Quantum} {Phase} {Transitions} with {Fermionic} {Neural} {Networks}},
	volume = {130},
	url = {https://link.aps.org/doi/10.1103/PhysRevLett.130.036401},
	doi = {10.1103/PhysRevLett.130.036401},
	number = {3},
	urldate = {2025-08-07},
	journal = {Phys. Rev. Lett.},
	author = {Cassella, Gino and Sutterud, Halvard and Azadi, Sam and Drummond, N. D. and Pfau, David and Spencer, James S. and Foulkes, W. M. C.},
	month = jan,
	year = {2023},
	note = {Publisher: American Physical Society},
	pages = {036401},
}

@article{pfa24,
	title = {Accurate computation of quantum excited states with neural networks},
	volume = {385},
	url = {https://www.science.org/doi/full/10.1126/science.adn0137},
	doi = {10.1126/science.adn0137},
	number = {6711},
	urldate = {2025-08-07},
	journal = {Science},
	author = {Pfau, David and Axelrod, Simon and Sutterud, Halvard and von Glehn, Ingrid and Spencer, James S.},
	month = aug,
	year = {2024},
	note = {Publisher: American Association for the Advancement of Science},
	pages = {eadn0137},
}

@misc{spe20,
      title={Better, Faster Fermionic Neural Networks}, 
      author={James S. Spencer and David Pfau and Aleksandar Botev and W. M. C. Foulkes},
      year={2020},
      eprint={2011.07125},
      archivePrefix={arXiv},
      primaryClass={physics.comp-ph},
      url={https://arxiv.org/abs/2011.07125}, 
}

@article{ben07,
  title={Scaling learning algorithms towards ai},
  author={LeCun, Yann and Bengio, Yoshua and others},
  journal={Large-scale kernel machines},
  volume={5},
  pages={127--168},
  year={2007},
  publisher={MIT Press},
  url={https://api.semanticscholar.org/CorpusID:15559637}
}

@Article{car17,
author={Carrasquilla, Juan
and Melko, Roger G.},
title={Machine learning phases of matter},
journal={Nature Physics},
year={2017},
month={May},
day={01},
volume={13},
number={5},
pages={431-434},
issn={1745-2481},
doi={10.1038/nphys4035},
url={https://doi.org/10.1038/nphys4035}
}

@article{car19,
  title = {Machine learning and the physical sciences},
  author = {Carleo, Giuseppe and Cirac, Ignacio and Cranmer, Kyle and Daudet, Laurent and Schuld, Maria and Tishby, Naftali and Vogt-Maranto, Leslie and Zdeborov\'a, Lenka},
  journal = {Rev. Mod. Phys.},
  volume = {91},
  issue = {4},
  pages = {045002},
  numpages = {39},
  year = {2019},
  month = {Dec},
  publisher = {American Physical Society},
  doi = {10.1103/RevModPhys.91.045002},
  url = {https://link.aps.org/doi/10.1103/RevModPhys.91.045002}
}

@article{alb14,
title = {The Inverse Ising Problem},
journal = {Physics Procedia},
volume = {57},
pages = {99-103},
year = {2014},
note = {Proceedings of the 27th Workshop on Computer Simulation Studies in Condensed Matter Physics (CSP2014)},
issn = {1875-3892},
doi = {https://doi.org/10.1016/j.phpro.2014.08.140},
url = {https://www.sciencedirect.com/science/article/pii/S1875389214002855},
author = {Joseph Albert and Robert H. Swendsen},
}

@article{gyo90,
  title = {First-order transition to perfect generalization in a neural network with binary synapses},
  author = {Gy\"orgyi, G\'eza},
  journal = {Phys. Rev. A},
  volume = {41},
  issue = {12},
  pages = {7097--7100},
  numpages = {0},
  year = {1990},
  month = {Jun},
  publisher = {American Physical Society},
  doi = {10.1103/PhysRevA.41.7097},
  url = {https://link.aps.org/doi/10.1103/PhysRevA.41.7097}
}

@article{gue18,
   author = "Guest, Dan and Cranmer, Kyle and Whiteson, Daniel",
   title = "Deep Learning and Its Application to LHC Physics", 
   journal= "Annual Review of Nuclear and Particle Science",
   year = "2018",
   volume = "68",
   number = "Volume 68, 2018",
   pages = "161-181",
   doi = "https://doi.org/10.1146/annurev-nucl-101917-021019",
   url = "https://www.annualreviews.org/content/journals/10.1146/annurev-nucl-101917-021019",
   publisher = "Annual Reviews",
   issn = "1545-4134",
   type = "Journal Article",
  }

@article{nta15,
doi = {10.1088/0004-637X/803/2/50},
url = {https://dx.doi.org/10.1088/0004-637X/803/2/50},
year = {2015},
month = {apr},
publisher = {The American Astronomical Society},
volume = {803},
number = {2},
pages = {50},
author = {M. Ntampaka and H. Trac and D. J. Sutherland and N. Battaglia and B. Póczos and J. Schneider},
title = {A MACHINE LEARNING APPROACH FOR DYNAMICAL MASS MEASUREMENTS OF GALAXY CLUSTERS},
journal = {The Astrophysical Journal},
}

@article{sar17,
  title = {Quantum Entanglement in Neural Network States},
  author = {Deng, Dong-Ling and Li, Xiaopeng and Das Sarma, S.},
  journal = {Phys. Rev. X},
  volume = {7},
  issue = {2},
  pages = {021021},
  numpages = {17},
  year = {2017},
  month = {May},
  publisher = {American Physical Society},
  doi = {10.1103/PhysRevX.7.021021},
  url = {https://link.aps.org/doi/10.1103/PhysRevX.7.021021}
}

@article{pfa20,
  title = {Ab initio solution of the many-electron Schr\"odinger equation with deep neural networks},
  author = {Pfau, David and Spencer, James S. and Matthews, Alexander G. D. G. and Foulkes, W. M. C.},
  journal = {Phys. Rev. Res.},
  volume = {2},
  issue = {3},
  pages = {033429},
  numpages = {20},
  year = {2020},
  month = {Sep},
  publisher = {American Physical Society},
  doi = {10.1103/PhysRevResearch.2.033429},
  url = {https://link.aps.org/doi/10.1103/PhysRevResearch.2.033429}
}

@article{zha19,
  title = {Solving Statistical Mechanics Using Variational Autoregressive Networks},
  author = {Wu, Dian and Wang, Lei and Zhang, Pan},
  journal = {Phys. Rev. Lett.},
  volume = {122},
  issue = {8},
  pages = {080602},
  numpages = {6},
  year = {2019},
  month = {Feb},
  publisher = {American Physical Society},
  doi = {10.1103/PhysRevLett.122.080602},
  url = {https://link.aps.org/doi/10.1103/PhysRevLett.122.080602}
}

@article{wen22,
doi = {10.1088/2752-5724/ac681d},
url = {https://dx.doi.org/10.1088/2752-5724/ac681d},
year = {2022},
month = {may},
publisher = {IOP Publishing},
volume = {1},
number = {2},
pages = {022601},
author = {Tongqi Wen and Linfeng Zhang and Han Wang and Weinan E and David J Srolovitz},
title = {Deep potentials for materials science},
journal = {Materials Futures},
}

@article{bcs57,
  title = {Theory of Superconductivity},
  author = {Bardeen, J. and Cooper, L. N. and Schrieffer, J. R.},
  journal = {Phys. Rev.},
  volume = {108},
  issue = {5},
  pages = {1175--1204},
  numpages = {0},
  year = {1957},
  month = {Dec},
  publisher = {American Physical Society},
  doi = {10.1103/PhysRev.108.1175},
  url = {https://link.aps.org/doi/10.1103/PhysRev.108.1175}
}

@article{and73,
title = {Resonating valence bonds: A new kind of insulator?},
journal = {Materials Research Bulletin},
volume = {8},
number = {2},
pages = {153-160},
year = {1973},
issn = {0025-5408},
doi = {https://doi.org/10.1016/0025-5408(73)90167-0},
url = {https://www.sciencedirect.com/science/article/pii/0025540873901670},
author = {P.W. Anderson},
}

@article{lau83,
  title = {Anomalous Quantum Hall Effect: An Incompressible Quantum Fluid with Fractionally Charged Excitations},
  author = {Laughlin, R. B.},
  journal = {Phys. Rev. Lett.},
  volume = {50},
  issue = {18},
  pages = {1395--1398},
  numpages = {0},
  year = {1983},
  month = {May},
  publisher = {American Physical Society},
  doi = {10.1103/PhysRevLett.50.1395},
  url = {https://link.aps.org/doi/10.1103/PhysRevLett.50.1395}
}

@article{cyb89,
author={Cybenko, G.},
title={Approximation by superpositions of a sigmoidal function},
journal={Mathematics of Control, Signals and Systems},
year={1989},
month={Dec},
day={01},
volume={2},
number={4},
pages={303-314},
issn={1435-568X},
doi={10.1007/BF02551274},
url={https://doi.org/10.1007/BF02551274}
}

@article{nag19,
	title = {Variational Quantum Monte Carlo Method with a Neural-Network Ansatz for Open Quantum Systems},
	volume = {122},
	issn = {0031-9007, 1079-7114},
	url = {https://link.aps.org/doi/10.1103/PhysRevLett.122.250501},
	doi = {10.1103/PhysRevLett.122.250501},
	pages = {250501},
    year={2019},
	number = {25},
	journal = {Physical Review Letters},
	shortjournal = {Phys. Rev. Lett.},
	author = {Nagy, Alexandra and Savona, Vincenzo},
	urldate = {2024-01-17},
	date = {2019-06-28},
	langid = {english},
}

@article{rot23,
	title = {High-accuracy variational Monte Carlo for frustrated magnets with deep neural networks},
	volume = {108},
	issn = {2469-9950, 2469-9969},
	url = {https://link.aps.org/doi/10.1103/PhysRevB.108.054410},
	doi = {10.1103/PhysRevB.108.054410},
	pages = {054410},
    year={2023},
	number = {5},
	journal = {Physical Review B},
	shortjournal = {Phys. Rev. B},
	author = {Roth, Christopher and Szabó, Attila and {MacDonald}, Allan H.},
	urldate = {2024-01-17},
	date = {2023-08-08},
	langid = {english},
}

@article{yan20,
	title = {Deep learning-enhanced variational Monte Carlo method for quantum many-body physics},
	volume = {2},
	issn = {2643-1564},
	url = {https://link.aps.org/doi/10.1103/PhysRevResearch.2.012039},
	doi = {10.1103/PhysRevResearch.2.012039},
	pages = {012039},
    year={2020},
	number = {1},
	journal = {Physical Review Research},
	shortjournal = {Phys. Rev. Research},
	author = {Yang, Li and Leng, Zhaoqi and Yu, Guangyuan and Patel, Ankit and Hu, Wen-Jun and Pu, Han},
	urldate = {2024-01-17},
	date = {2020-02-14},
	langid = {english},
}

@article{qia22,
	title = {Interatomic force from neural network based variational quantum Monte Carlo},
	volume = {157},
	issn = {0021-9606, 1089-7690},
	url = {https://pubs.aip.org/jcp/article/157/16/164104/2841958/Interatomic-force-from-neural-network-based},
	doi = {10.1063/5.0112344},
	pages = {164104},
	number = {16},
	journal = {The Journal of Chemical Physics},
	author = {Qian, Yubing and Fu, Weizhong and Ren, Weiluo and Chen, Ji},
	urldate = {2024-01-17},
	date = {2022-10-28},
	langid = {english},
    year = {2022}
}

@article{car21,
  title = {Neural-network quantum states for periodic systems in continuous space},
  author = {Pescia, Gabriel and Han, Jiequn and Lovato, Alessandro and Lu, Jianfeng and Carleo, Giuseppe},
  journal = {Phys. Rev. Research},
  volume = {4},
  issue = {2},
  pages = {023138},
  numpages = {8},
  year = {2022},
  month = {5},
  publisher = {American Physical Society},
  doi = {10.1103/PhysRevResearch.4.023138},
  url = {https://link.aps.org/doi/10.1103/PhysRevResearch.4.023138}
}

@article{fre25,
  title = {Neural network based nodal structure optimization for interacting fermionic systems},
  author = {Freitas, William and Abreu, B. and Vitiello, S. A.},
  journal = {Phys. Rev. B},
  volume = {112},
  issue = {16},
  pages = {165109},
  numpages = {9},
  year = {2025},
  month = {Oct},
  publisher = {American Physical Society},
  doi = {10.1103/x6cj-b6lj},
  url = {https://link.aps.org/doi/10.1103/x6cj-b6lj}
}

@article{fre24,
author={Freitas, William
and Abreu, Bruno
and Vitiello, S. A.},
title={Modeling $^4${He}$_{N}$ Clusters with Wave Functions Based on Neural Networks},
journal={Journal of Low Temperature Physics},
year={2024},
month={Mar},
day={06},
issn={1573-7357},
doi={10.1007/s10909-024-03061-w},
url={https://doi.org/10.1007/s10909-024-03061-w}
}

@article{fre23,
  doi = {10.22331/q-2023-12-18-1209},
  url = {https://doi.org/10.22331/q-2023-12-18-1209},
  title = {Synergy between deep neural networks and the variational {M}onte {C}arlo method for small {$^4He_N$} clusters},
  author = {Freitas, William and Vitiello, S. A.},
  journal = {{Quantum}},
  issn = {2521-327X},
  publisher = {{Verein zur F{\"{o}}rderung des Open Access Publizierens in den Quantenwissenschaften}},
  volume = {7},
  pages = {1209},
  month = {dec},
  year = {2023}
}

@article{rum86,
author={Rumelhart, David E.
and Hinton, Geoffrey E.
and Williams, Ronald J.},
title={Learning representations by back-propagating errors},
journal={Nature},
year={1986},
month={Oct},
day={01},
volume={323},
number={6088},
pages={533-536},
issn={1476-4687},
doi={10.1038/323533a0},
url={https://doi.org/10.1038/323533a0}
}

@book{hinton,
    author = {Rumelhart, David E. and McClelland, James L. and PDP Research Group},
    title = "{Parallel Distributed Processing, Volume 1: Explorations in the Microstructure of Cognition: Foundations}",
    publisher = {The MIT Press},
    year = {1986},
    month = {07},
    isbn = {9780262291408},
    doi = {10.7551/mitpress/5236.001.0001},
    url = {https://doi.org/10.7551/mitpress/5236.001.0001},
}

@book{goodfellow,
  title={Deep learning},
  author={Goodfellow, Ian and Bengio, Yoshua and Courville, Aaron},
  year={2016},
  publisher={MIT press}
}

@Article{mcc43,
author={McCulloch, Warren S.
and Pitts, Walter},
title={A logical calculus of the ideas immanent in nervous activity},
journal={The bulletin of mathematical biophysics},
year={1943},
month={Dec},
day={01},
volume={5},
number={4},
pages={115-133},
issn={1522-9602},
doi={10.1007/BF02478259},
url={https://doi.org/10.1007/BF02478259}
}

@article{owid-moores-law,
    author = {Max Roser and Hannah Ritchie and Edouard Mathieu},
    title = {What is Moore's Law?},
    journal = {Our World in Data},
    year = {2023},
    note = {https://ourworldindata.org/moores-law}
}

@article{moo98,
Author = {Moore, GE},
Title = {Cramming more components onto integrated circuits (Reprinted from
   Electronics, pg 114-117, April 19, 1965)},
Journal = {PROCEEDINGS OF THE IEEE},
Year = {1998},
Volume = {86},
Number = {1},
Pages = {82-85},
Month = {JAN},
DOI = {10.1109/JPROC.1998.658762},
ISSN = {0018-9219},
Unique-ID = {WOS:000071717000014},
}

@article{wan23,
Author = {Wang, Zhangu and Zhan, Jun and Duan, Chunguang and Guan, Xin and Lu,
   Pingping and Yang, Kai},
Title = {A Review of Vehicle Detection Techniques for Intelligent Vehicles},
Journal = {IEEE TRANSACTIONS ON NEURAL NETWORKS AND LEARNING SYSTEMS},
Year = {2023},
Volume = {34},
Number = {8},
Pages = {3811-3831},
Month = {AUG},
url = {https://doi.org/10.1109/TNNLS.2021.3128968},
}

@article{rau21,
Author = {Raucci, Umberto and Valentini, Alessio and Pieri, Elisa and Weir, Hayley
   and Seritan, Stefan and Martinez, Todd J.},
Title = {Voice-controlled quantum chemistry},
Journal = {NATURE COMPUTATIONAL SCIENCE},
Year = {2021},
Volume = {1},
Number = {1},
Pages = {42-45},
Month = {JAN},
url = {https://doi.org/10.1038/s43588-020-00012-9},
}

@article{hol18,
  title = {Nonlinear Network Description for Many-Body Quantum Systems in Continuous Space},
  author = {Ruggeri, Michele and Moroni, Saverio and Holzmann, Markus},
  journal = {Phys. Rev. Lett.},
  volume = {120},
  issue = {20},
  pages = {205302},
  numpages = {6},
  year = {2018},
  month = {May},
  publisher = {American Physical Society},
  doi = {10.1103/PhysRevLett.120.205302},
  url = {https://link.aps.org/doi/10.1103/PhysRevLett.120.205302}
}

@article{lei22,
Author = {Li, Lianhui and Lei, Bingbing and Mao, Chunlei},
Title = {Digital twin in smart manufacturing},
Journal = {JOURNAL OF INDUSTRIAL INFORMATION INTEGRATION},
Year = {2022},
Volume = {26},
Month = {MAR},
url = {https://doi.org/10.1016/j.jii.2021.100289},
}

@article{uch21,
  author       = {Adaku Uchendu and
                  Zeyu Ma and
                  Thai Le and
                  Rui Zhang and
                  Dongwon Lee},
  title        = {{TURINGBENCH:} {A} Benchmark Environment for Turing Test in the Age
                  of Neural Text Generation},
  journal      = {CoRR},
  volume       = {abs/2109.13296},
  year         = {2021},
  url          = {https://arxiv.org/abs/2109.13296},
}

@article{fuk75,
author={Fukushima, Kunihiko},
title={Cognitron: A self-organizing multilayered neural network},
journal={Biological Cybernetics},
year={1975},
month={Sep},
day={01},
volume={20},
number={3},
pages={121-136},
issn={1432-0770},
doi={10.1007/BF00342633},
url={https://doi.org/10.1007/BF00342633}
}

@article{sprng,
author = {Mascagni, Michael and Srinivasan, Ashok},
title = {Algorithm 806: SPRNG: a scalable library for pseudorandom number generation},
year = {2000},
issue_date = {Sept. 2000},
publisher = {Association for Computing Machinery},
address = {New York, NY, USA},
volume = {26},
number = {3},
issn = {0098-3500},
url = {https://doi.org/10.1145/358407.358427},
doi = {10.1145/358407.358427},
journal = {ACM Trans. Math. Softw.},
month = sep,
pages = {436–461},
numpages = {26}
}

@article{sch96,
    author = {Kosztin, Ioan and Faber, Byron and Schulten, Klaus},
    title = {Introduction to the diffusion Monte Carlo method},
    journal = {American Journal of Physics},
    volume = {64},
    number = {5},
    pages = {633-644},
    year = {1996},
    month = {05},
    issn = {0002-9505},
    doi = {10.1119/1.18168},
    url = {https://doi.org/10.1119/1.18168},
}

@article{rod21,
title = {Complete analytical solution to the quantum Yukawa potential},
journal = {Physics Letters B},
volume = {816},
pages = {136218},
year = {2021},
issn = {0370-2693},
doi = {https://doi.org/10.1016/j.physletb.2021.136218},
url = {https://www.sciencedirect.com/science/article/pii/S0370269321001581},
author = {M. Napsuciale and S. Rodríguez}
}

@misc{annvmc-github-alaias,
  author = {X. Author},
  title = {{Introduction to the artificial neural network-based variational
Monte Carlo program}},
  url = {https://some-website.com/annvmc-intro},
  year = {2025},
}

@misc{fre26sm,
  author = {William Freitas},
  title = {{Supplemental Material for Introduction to the artificial neural network-based variational Monte Carlo method}},
  url = {https://},
  year = {2026},
}

@misc{annvmc-github,
  author = {William Freitas},
  title = {{Artificial Neural Network-based Variational Monte Carlo Introduction}},
  url = {https://github.com/freitas-esw/annvmc-intro},
  year = {2025},
}

@misc{jaxgithub,
  author = {James Bradbury and Roy Frostig and Peter Hawkins and Matthew James Johnson and Chris Leary and Dougal Maclaurin and George Necula and Adam Paszke and Jake Vander{P}las and Skye Wanderman-{M}ilne and Qiao Zhang},
  title = {{JAX}: composable transformations of {P}ython+{N}um{P}y programs},
  url = {http://github.com/jax-ml/jax},
  version = {0.3.13},
  year = {2018},
}

@book{cohen,
  title={Quantum Mechanics (2 vol. set)},
  author={Cohen-Tannoudji, Claude and Diu, Bernard and Laloe, Frank and Dui, Bernard},
  year={2006},
  publisher={Wiley-Interscience}
}

@article{met53,
    author = {Metropolis, Nicholas and Rosenbluth, Arianna W. and Rosenbluth, Marshall N. and Teller, Augusta H. and Teller, Edward},
    title = "{Equation of State Calculations by Fast Computing Machines}",
    journal = {The Journal of Chemical Physics},
    volume = {21},
    number = {6},
    pages = {1087-1092},
    year = {1953},
    month = {06},
    issn = {0021-9606},
    doi = {10.1063/1.1699114},
    url = {https://doi.org/10.1063/1.1699114},
}

@article{ros58,
  title={The perceptron: a probabilistic model for information storage and organization in the brain.},
  author={Rosenblatt, Frank},
  journal={Psychological review},
  volume={65},
  number={6},
  pages={386},
  year={1958},
  publisher={American Psychological Association},
  url={https://psycnet.apa.org/doi/10.1037/h0042519}
}

@misc{adam,
      title={Adam: A Method for Stochastic Optimization}, 
      author={Diederik P. Kingma and Jimmy Ba},
      year={2017},
      eprint={1412.6980},
      archivePrefix={arXiv},
      primaryClass={cs.LG},
      url={https://arxiv.org/abs/1412.6980}, 
}

@misc{deepseek,
      title={DeepSeek-V3 Technical Report}, 
      author={DeepSeek-AI},
      year={2025},
      eprint={2412.19437},
      archivePrefix={arXiv},
      primaryClass={cs.CL},
      url={https://arxiv.org/abs/2412.19437}, 
}

@misc{bub25,
      title={Early science acceleration experiments with GPT-5}, 
      author={Sébastien Bubeck and Christian Coester and Ronen Eldan and Timothy Gowers and Yin Tat Lee and Alexandru Lupsasca and Mehtaab Sawhney and Robert Scherrer and Mark Sellke and Brian K. Spears and Derya Unutmaz and Kevin Weil and Steven Yin and Nikita Zhivotovskiy},
      year={2025},
      eprint={2511.16072},
      archivePrefix={arXiv},
      primaryClass={cs.CL},
      url={https://arxiv.org/abs/2511.16072}, 
}

@article{neumann,
  title={First Draft of a Report on the EDVAC},
  author={Von Neumann, John},
  journal={IEEE Annals of the History of Computing},
  volume={15},
  number={4},
  pages={27--75},
  year={1993},
  publisher={IEEE}
}

@incollection{eniac,
  title={The electronic numerical integrator and computer (eniac)},
  author={Goldstine, Herman H and Goldstine, Adele},
  booktitle={The Origins of Digital Computers: Selected Papers},
  pages={359--373},
  year={1946},
  publisher={Springer}
}

@incollection{randell,
title = {The COLOSSUS},
editor = {N. METROPOLIS and J. HOWLETT and GIAN-CARLO ROTA},
booktitle = {A History of Computing in the Twentieth Century},
publisher = {Academic Press},
address = {San Diego},
pages = {47-92},
year = {1980},
isbn = {978-0-12-491650-0},
doi = {https://doi.org/10.1016/B978-0-12-491650-0.50013-7},
url = {https://www.sciencedirect.com/science/article/pii/B9780124916500500137},
author = {B. RANDELL},
}

@book{welchman,
  title={The Hut Six Story: Breaking the Enigma Codes},
  author={Welchman, G.},
  isbn={9780070691803},
  lccn={lc81013657},
  year={1982},
  publisher={McGraw-Hill}
}

@book{goldstine,
  title={The computer from Pascal to von Neumann},
  author={Goldstine, Herman H},
  year={1993},
  publisher={Princeton University Press}
}

@article{lov1842,
  title={Sketch of the analytical engine invented by Charles Babbage, by LF Menabrea, officer of the military engineers, with notes upon the memoir by the translator},
  author={Lovelace, Augusta Ada},
  journal={Taylor’s Scientific Memoirs},
  volume={3},
  pages={666--731},
  year={1842},
  url={https://www.fourmilab.ch/babbage/sketch.html}
}

\appendix
\newpage

\section*{Supplemental Material for: ``\textit{Introduction to the artificial neural network-based variational Monte Carlo method}''}

\section{Historical background}

The desire to create AI has been spread throughout humanity at least since ancient Greece. Some old legends tell about mythical figures such as Hephaestus, god of artisans, blacksmiths, fire and metallurgy. According to Greek mythology, Pandora, the first mortal woman, was forged by Hephaestus, revealing the existing thoughts of that time about human intelligence created artificially. Moreover, this myth also mentions that virtues such as grace and persuasion were given to Pandora by the gods. The current computers can be seen from a similar aspect, where a sequence of adaptable instructions with a specific purpose, in other words, an algorithm is given to the machine in order to automate a procedure. The programmable essence of computers is alike the virtues offered to Pandora and the aspiration to create computers that think still pondering in society nowadays, as they did in ancient Greece. 

However, to reach the current stage of AI technologies, important developments were necessary, both intellectually and in terms of technological innovation. In particular, the beginning of programmable computers was due to the desire to automate numerical calculations. Historically, one of the forerunners of this desire was Galileo Galilei with his mathematical formulation of the natural sciences \cite{goldstine}. By uniting natural sciences and math, a demand for a machine that can calculate arose among scientists. That influenced Pascal in the creation of the addition machine in 1642 and, afterwards, it influenced Leibniz in the invention of the Leibniz wheel in 1673. Pascal's device could sum and subtract, while Leibniz's device could also make products and divisions. Later, in the 19th century, those devices supported Charles Babbage on the development of both difference and analytical engines, that are considered prototypes of programmable computers. Contemporaneously and based on Babbage's work, Ada Lovelace proposed a method represented by a sequence of steps to compute the Bernoulli numbers through the analytical engine \cite{lov1842}. Frequently, this method is referenced as the first algorithm created to be processed by a machine, which made Ada known as the first programmer.

In the following century, World War II had an important influence on the development of such machines, leading to the creation of the Bombe, an electro-mechanical device specially assembled to decipher German messages encrypted by the ENIGMA. Among the most important names in this project were Alan Turing due to the introductory design and Gordon Welchman for a crucial refinement on the project \cite{welchman}. Simultaneously, Turing also aided the development of the computers Colossus Mark 1 and Colossus Mark 2, designed by Tommy Flowers and influenced by Max Newman's work \cite{randell}. These computers also aimed to decipher other kinds of encrypted messages, however, they were based on thermionic valves to perform operations and were more versatile, they could be programmed using switches and plugs. 

Not much later than that, in 1945, the machine regarded as the first fully electronic, general-purpose computer, the ENIAC was completed \cite{eniac}. It operated in the decimal base, and the design is attributed to John Mauchly and John Eckert. Also, both of them worked together with John von Neumann on the successor of ENIAC, the EDVAC computer, which operated on the binary base and was designed to be a stored-program computer, in other words, the program instruction and data could be stored in an accessible memory \cite{neumann}. That still is a key feature of modern computers and made programming rather simpler compared with ENIAC, in which one has to program through plugboards and switches.

In this context, perhaps also influenced by the same desires as the ancient Greeks, Alan Turing proposed the imitation game \cite{turing50}, which aims to probe if machines are capable of thinking. However, due to the philosophical nature of the meaning of thinking, the question has to be replaced by a feasible experiment. In that sense, the aim of this experiment becomes to determine if machines can mimic human intelligence. Therefore, the Turing test entails three participants: one human interviewer and two interviewed participants, one of them a human and the other a machine. Then, the interviewer talks separately with each interviewed through text messages. Their aim is to find who is the machine and who is the human. Consequently, the machine will try to mimic the behaviour of the human in the conversation and, hence, deceive the interviewer. If the interviewer mistakes the machine for a human, then this machine is capable of thinking in the Turing test sense. 

With the development of large generative models such as GPT-5 \cite{bub25} and DeepSeek \cite{deepseek}, the Turing test is on the verge of being accomplished in the way it was constructed \cite{uch21}. 
Not only that, but many other useful technologies such as augmented and virtual reality, autonomous vehicles \cite{wan23}, virtual assistants \cite{rau21}, and digital twins \cite{lei22} employ AI tools. However, to reach the current AI state-of-the-art technologies, from the 50s to nowadays, several important improvements both intellectually and technologically were necessary.

Regarding the technological advancements, one of the most important factors was the miniaturisation of computer components. From the beginning of commercially available general-purpose computers such as the Ferranti Mark I to current AMD Ryzen Threadripper processors, the size of computers has reduced dramatically. Two important events in this period of time were the release of the first commercially available microprocessor by Intel in 1971 and the IBM personal computers in 1981. At that time, the number of transistors in an integrated circuit was about $10^5$. Even before that, a phenomenon known as Moore's law \cite{moo98} was observed. The law states that the number of transistors in an integrated circuit approximately doubles every other year. This tendency of growing in numbers is also supported by the miniaturisation, in other words, they also increase the density to keep the devices at a reasonable size. The trend persists even nowadays, $10^{11}$ is the current order of transistors in a microchip \cite{owid-moores-law}.

On the other hand, and parallel to the technological advancements, theoretical progress in the AI field were also necessary. 
A pivotal tool in this process is deep learning. However, there were three great ``waves'' of development for this tool to achieve its current design \cite{goodfellow}. The first ``wave'' is known as \textbf{cybernetics} and had its beginning during the 40s, while its peak of popularity was around the end of the 60s. Throughout this period of time, the forerunners of deep learning models were only linear models, as, for example, the neuron of McCulloch-Pitts \cite{mcc43}. This model was able to distinguish between two categories of information. However, in order for the neuron to correctly predict between the two categories, it was necessary to adjust the model parameters manually. A few years later, another linear model known as \textit{perceptron} \cite{ros58} was proposed, and, by learning from examples, it was capable of determining in which of the two categories the information belonged. Currently, the perceptron is still being broadly used in the context of machine learning and also as components of models in deep learning applications.

The second ``wave'' of development became historically known as \textbf{connectionism} and had its beginning during the 80s. The peak of the wave was around the mid-90s, and several advancements took place in this period of time, with many of them remaining up to nowadays as basic blocks for the construction of modern ANNs. Among the core tools that connectionism popularised, one important contribution was the model called \textit{cognitron} \cite{fuk75}, which introduced non-linear ways to process the data given to the model. Before that, only one basic unit, or neuron, was used to compose most of the models. However, the central idea behind the connectionism was to reach intelligent behaviour through connecting computational units together, inspired by the biological neurons. Therefore, it was at this point in time that the forerunner models of the modern ANNs were proposed.

In this context, the concept of distributed representation \cite{hinton} emerged. In order to understand this idea, consider a vision system that is capable of recognizing shelves, desks, and chairs, and any of them might be made from wood, metal, or plastic. One way of organizing the neurons of a model to perform the recognition of the possible objects is to assign one neuron for each combination of furniture and material. Another way would be to split the neurons into two layers of three neurons each, where one layer can separate the type of furniture, while the other would classify the material. By connecting densely the two layers, only six neurons would be necessary to perform the recognition in this example. That would be the distributed representation of the model, which reduces the number of neurons to achieve the desired goal. This kind of representation enables the neurons to understand complex concepts in terms of simpler concepts. Another remarkable contribution was the success in employing the back-propagation algorithm \cite{rum86} to train the models and its popularization. Currently, this algorithm is still one of the core tools when training models.

From 2006 on, the third ``wave'' begins with one work demonstrating that a specific model can be trained efficiently with a strategy called greedy layer-by-layer pre-training \cite{hil07}. The following works showed that the same strategy could be used to train other types of models and systematically improve their performance \cite{ben07}. It was in this wave of works that the term ``deep learning'' was popularized. Eventually, the advancements of such techniques reached physics in all its several domains, where the first application was in condensed matter physics \cite{car17}. More recent applications are mentioned in the introduction.

Although the extensive use of machine learning tools applied to physics is recent, there has been cross-breeding between statistical physics and machine learning since connectionism's wave \cite{car19}. An interesting example is the first-order phase transition that takes place in perceptrons when they are being trained by examples \cite{gyo90}. As the number of examples increases, the model reaches the reference state. Another notable connection between these two fields is the Boltzmann machines, which are known as the inverse Ising problem \cite{alb14} within the physics literature. In this problem, the task is to determine the coupling parameters between spins based on a dataset of Ising model configurations. Another key aspect that both fields share is a special interest in the optimization problem. Particularly, in the context of quantum mechanics, the minimization of the energy functional, also known as the VM, has a direct analogy with machine learning.

\section{Variational Monte Carlo method}

Given a normalizable trial wave function $\psi_\theta$ parametrized by variational parameters $\theta$, the energy functional $E[\theta]$ yields an energy expectation value that is an upper bound for the GS energy, namely

\begin{equation}\label{eq:exp-h}
E[\theta] = 
\frac
  {\int d\vct x \ \psi_\theta^*(\vct x)\mathcal H \psi_\theta(\vct x)}
  {\int d\vct x \ |\psi_\theta(\vct x)|^2 } \ge
E_0 ~.
\end{equation}

\noindent
This also means that the expectation value $E[\theta]$ has a global minimum that is larger or equal to the GS energy. Minimizing the energy functional with respect to the parameters identifies the function that provides the best approximation to the GS within the variational family $\psi_\theta$. Moreover, considering the scope of this work, the Hamiltonian will be written simply as the sum of kinetic energy and potential energy $\mathcal H = \hat K + V(\vct x)$.

Several interesting systems can be studied throughout trial wave functions. However, the high dimensionality of the integrals involved in \cref{eq:exp-h} often makes the analytical computation prohibitive. The variational Monte Carlo (VMC) method is the key to overcome this limitation. The formulation of the method requires to rewrite the energy expectation value in terms of the probability density function (PDF)
$
p ( \vct x ) = \nf
	{ | \psi_\theta (\vct x)|^2 }
	{ \mathcal N } \ ,
$
where the normalization factor is given by
$
\mathcal N = \int d\vct x \ | \psi_\theta (\vct x)|^2 ~,
$
and in terms of the local energy function, which is written as 
$
 E_L(\vct x) = \psi_\theta^{-1} (\vct x)~ \mathcal H~ \psi_\theta(\vct x) ~.
$
By doing so, 
the energy expectation value reads

\begin{equation}
E[\theta] = 
\frac
  {\int d\vct x \ |\psi_\theta(\vct x) |^2
  \frac {\mathcal H \psi_\theta(\vct x)}{\psi_\theta(\vct x)}}
  {\int d\vct x \ |\psi_\theta(\vct x)|^2 } =
\int d\vct x \ p(\vct x) E_L (\vct x) 
~.
\end{equation}

\noindent
In principle, since $p(\vct x)$ is a PDF, the integral can be viewed as a continuous weighted sum of the local energy over the configuration space. However, by drawing a set of $M$ samples $\{\vct x_i\}_{i=1}^M$ distributed according to $p(\vct x)$, then, the integral could be estimated as a simple average $E_{\rm VMC} \simeq \sum_{i=1}^M \nf {E_L (\vct x_i)} M $, which converges to the exact value in the limit of large $M$, with a statistical error that scales as $\nf 1 {\sqrt M}$. That procedure is called Monte Carlo integration.

Nevertheless, exactly sampling $p(\vct x)$ can be as hard as solving the full Schroedinger equation. Therefore, a frequent approach when performing a Monte Carlo integration is to employ the Metropolis algorithm~\cite{met53}. This algorithm is a advanced sampling method that uses rejection techniques, in other words, it involves explicitly proposing tentative values and accepting or rejecting based upon some rules. The method is also iterative in the sense that it utilizes the current values to propose trial ones. The procedure is summarized as follows: 

\begin{enumerate}[i]
    \item chose a initial configuration $\vct x_0$ and set $j=0$;
    \item propose a trial move according to $\vct x_t=\vct x_j + \Delta (\nf 1 2 - \bm\xi)$;
    \item compute the density probability ratio $q=\nf {p(\vct x_t)}{p(\vct x_i)}$;
    \item if $q>\xi'$ accept the trial move $\vct x_{j+1}=\vct x_t$ or else reject it $\vct x_{j+1}=\vct x_j$;
    \item set $j=j+1$ and return to step ii).
\end{enumerate}

Note that the acceptation probability $q=\nf{|\psi_\theta(\vct x_{t})|^2}{|\psi_\theta(\vct x_j)|^2}$ and, therefore, the procedure does not depend on the normalization $\mathcal N$. Moreover, $\bm\xi$ and $\xi'$ denote, respectively, a $ND$-dimensional vector and a scalar drawn from a uniform distribution on the interval $[0,1)$. Generating pseudo random numbers uniformly distributed is a field of its own, where pseudo random number generators such as SPRNG~\cite{sprng} and JAX~\cite{jaxgithub} are examples of efficient generators. Lastly, the step size $\Delta$ controls the acceptance ratio, that is, how many moves are accepted among a certain number of steps. The acceptance is large for small step sizes, while the opposite behaviour happens when the step size is large. For the described version of the Metropolis algorithm, the most efficient acceptation rate requires adjusting the step size in order to reach 50$\%$ acceptance. 

In this work, the initial configurations are chosen randomly. Its important to point out that the initial steps are highly correlated to the initial configurations and, hence, are not sampled according to $p(\vct x)$. Therefore, the energy estimation cannot be trusted in those steps.
Nevertheless, after the initial steps, that should be discarded, the procedure converges towards the equilibrium, namely, to the correct distribution. Then, the energy estimation is performed using the correctly sampled configurations as described before. Since this method is stochastic and subsequent steps are correlated, the unbiased estimator of the standard error of the mean (SEM) together with the blocking procedure is necessary in order to evaluate the statistical error present in the estimated result. More details can be found on~\cref{ap:sem}.

Finally, aiming to minimize the energy functional $E[\theta]$, the gradient with respect to the variational parameters is required. By carefully computing the gradient of the ratio present in \cref{eq:exp-h} and assuming the trial wave function is real, the expression reads

\begin{equation}
\nabla_\theta E[\theta] = 
 \int d\vct x\ p(\vct x)\left(E_L(\vct x) - 
 \int d\vct x' p(\vct x')E_L(\vct x')\right) 
 \nabla_\theta \ln |\psi_\theta(\vct x)|^2 ~.
\end{equation}

\noindent
Employing the same techniques as before, the estimation of this integral is also done through Monte Carlo integration. Accordingly, the gradient is estimated by the average $\nabla_\theta E \simeq M^{-1} \sum_{i=1}^M {\left[E_L(\vct x_i)-E_{\rm VMC}\right]\nabla_\theta\ln|\psi_\theta(\vct x_i)|^2}$, where $\vct x_i$ are the same configurations sampled from $p(\vct x)$ used to estimate $E_{\rm VMC}$. Once these quantities are estimated, minimization methods such as the gradient descent might be applied in order to find the minimum of the energy functional and, consequently, the best approximation for the GS within the class of trial functions $\psi_\theta$. 

\section{Blocking method and standard error estimation}\label{ap:sem}

Since the Monte Carlo integration is based on a Markov chain, successive samples are correlated. Therefore, the effective number of samples is smaller than the actual number of samples $M$. Consequently, the usual standard error estimation is too optimistic. To illustrate the procedure, consider a series of data $E_1$, $E_2$, ..., $E_M$. The blocking procedure consists on grouping the data in blocks of size $b$ and computing their averages,

\begin{equation}
\bar E^{(b)}_k =  \frac 1 b \sum_{i=(k-1)b+1}^{kb} E_i ~.
\end{equation}

\noindent
The variance of the blocked-averaged data is also computed through

\begin{equation}
\sigma_b^2 = \frac 1 {M-1} \sum_i^{\nf M b} 
\left( 
\bar E^{(b)}_i - \bar E
\right)^2 ~,
\end{equation}

\noindent where the total average data $\bar E$ is computed over all data set, namely, $\bar E = \sum_i^M \nf {E_i} M$. Finally, the standard error of the mean is given by

\begin{equation}
{\rm SEM} = \sqrt{\frac {\sigma_b^2} M}~.
\end{equation}

For small block sizes $b$, the SEM is underestimated. As the block size is gradually increased, eventually, the error saturates and start fluctuating around a constant value. In this later stage, the blocks are large enough to make the blocked-averaged data $\bar E^{(b)}_i$ to be effectively independent. As a consequence, the correct error estimation is given in this length range of block sizes.

\section{Adaptive Moment Estimation optimizer}\label{ap:adam}

When dealing with noisy gradients coming from stochastic methods and loss landscapes in high dimensional spaces, naive gradient-based optimization methods are likely to perform poorly. The adaptive moment estimation~\cite{adam} is a popular optimization algorithm used in machine learning in order to overcome the mentioned limitations of simple gradient descent methods. The algorithm employs weighted moving averages to estimate first and second moments of the loss gradient, which are used in the update of the model parameters. The procedure aids in the control of the step length individually for each parameter and address convergence issues.

Denoting the loss gradient with respect to the parameters as $g_t = \nabla\mathcal L [\theta_t]$, the algorithm updates iteratively state variables $p_t$ and $q_t$ according to the following equations:

\begin{subequations}
\begin{align}
p_t &=  \beta_1 ~p_{t-1} + (1-\beta_1) ~g_t~, \\
q_t &= \beta_1 ~q_{t-1} + (1-\beta_2) ~g_t^2 ~,\\
\hat p_t &= \frac {p_t}{1-\beta_1^t}~, \\
\hat q_t &= \frac {q_t}{1-\beta_2^t}~,
\end{align}
\end{subequations}

\noindent where $p_0=q_0=0$ and the hyperparameters $\eta$, $\beta_1$, $\beta_2$, and $\epsilon$ controls the performance of the algorithm. Typical values are $\eta=10^{-3}$, $\beta_1=0.9$, $\beta_2=0.999$, and $\epsilon=10^{-8}$. Furthermore, the proposed update for the parameters is written as:

\begin{equation}
\Delta\theta_t = - \frac {\eta~\hat p_t}{\sqrt{\hat q_t} + \epsilon}~.
\end{equation}

\noindent Therefore, the parameters are updated according to
$\theta_{t+1} = \theta_{t} + \Delta\theta_{t}$. Additionally, the learning rate $\eta$ can also be adapted as a function of $t$, granting that the steps become smaller as the optimization steps $t$ increase. One example of such learning rate schedule would be 

\begin{equation}
\eta(t) = \frac{\eta_0}{\left(1+\nf t {a_{\rm delay}}\right)^{a_{\rm decay}}}
\end{equation}

\noindent
where $\eta_0$ is the initial learning rate, $a_{\rm delay}$ controls from which step the suppression of the learning rate will start, and $a_{\rm decay}$ controls how fast is the decay.

\section{VMC algorithm}\label{ap:algs}

Here is presented pseudo code for the implementation details of the VMC algorithm. The functions ${\rm RandU}(\cdot)$ and ${\rm RandN}(\cdot)$ are supposed to be pseudo random number generators for the uniform and normal distribution, respectively. The expected input is the length of the dimensions of the array to be generated, that is, the size of the array if it is a vector or the number of rows and columns if it is a matrix. 

On the other hand, the implementation of the optimization procedure is presented in~\cref{alg:vmc}. 
The user defined parameters are the number of optimization iterations denoted by $N_{OI}$, and the batch size (or number of walkers) denoted by $M$. Usually, the quantum Monte Carlo codes are written in the logarithm domain in order to avoid loss of numerical precision. Note that, however, the pseudo code is not taking that detail into account, although the available code~\cite{annvmc-github-alaias} does that. Therefore, for more details, the reader is encouraged to look carefully into the code employed in the simulations.

\begin{algorithm}[tb]
\caption{NN-VMC algorithm}\label{alg:vmc}
\begin{algorithmic}[1]
\Require $M, N_{\rm OI}, n_0,\theta$
\State $\vct x_0 \gets {\rm RandU}(n_0)$
\For{$i \gets$ 1 \textbf{to} $N_{OI}$}
\For{$j \gets$ 1 \textbf{to} $M$}
\State $\bm\xi \gets {\rm RandU}(n_0)$
\State $\vct x_t \gets \vct x_j + \Delta(\nf 1 2 - \bm\xi)$
\State $q \gets p(\vct x_t)/p(\vct x_j)$
\State $\xi' \gets {\rm RandU}(1)$
\If{$q>\xi'$}
\State $\vct x_j \gets \vct x_t$
\EndIf
\State $E_j \gets E_L(\vct x_j)$
\State $g_{j,\theta} \gets 2\nabla_\theta\ln|\psi_\theta(\vct x_j)|$
\EndFor
\State $E_{\rm VMC} \gets \langle E_j \rangle_{j=1}^M$
\State $\Delta E_{\rm VMC} \gets {\rm SEM}[E_j]$
\State $G_\theta \gets \left\langle(E_j-E_{\rm VMC})~g_{j,\theta}\right\rangle_{j=1}^M$
\State $\Delta\theta \gets {\rm ADAM}[G_\theta]$
\State $\theta \gets \theta + \Delta\theta$
\EndFor
\end{algorithmic}
\end{algorithm}


\section{Extra examples}

\subsection{Harmonic oscillator}

As one of the simplest and solvable systems, a particle in a one dimensional harmonic oscillator is one of the best proving grounds for testing the ANN based trial wave functions. For examples like this that are described in terms of only one variable, that is, $N=1$ and $D=1$, the system component is written as $x\equiv\vct x$. Considering a particle of mass $m$ and a confinement frequency of $\omega$, the potential $V(x)$ reads
\begin{equation}
V(x) = \frac {m\omega^2x^2} 2 ~.
\end{equation}
%
Choosing $\hbar\omega$ as the energy units and $a_{\rm ho}=\sqrt{\nf \hbar{m\omega}}$ as length units, the exact GS energy is $E_0=\nf 1 2$ and the dimensionless wave function is given by $\psi(\hat x)=\pi^{-\nf 1 4}\exp\left(-\nf{\hat x^2} 2\right)$.

\begin{figure}[tb]
    \centering
    \includegraphics[width=0.85\linewidth]{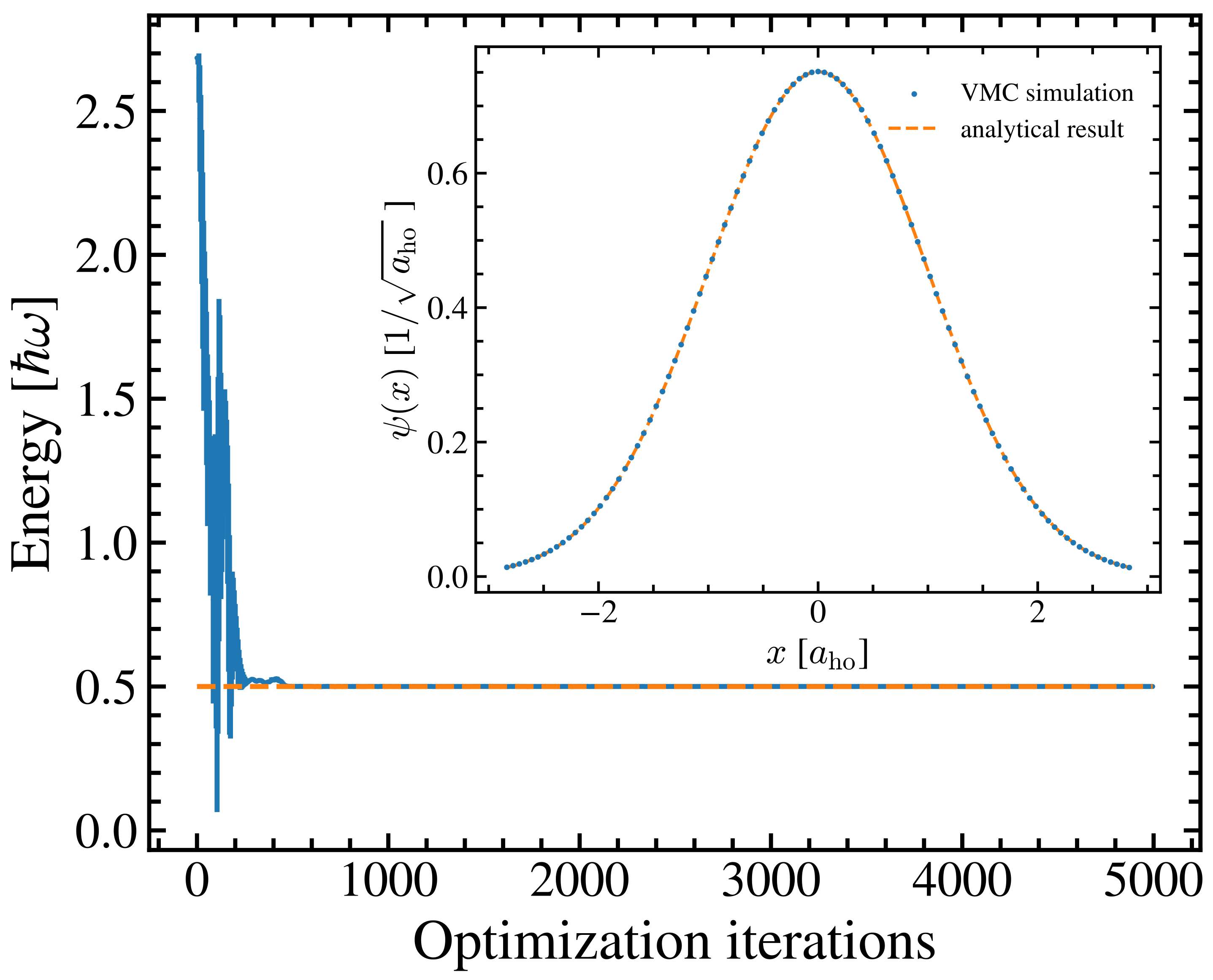}
    \caption{Minimization process for the harmonic oscillator potential showing the energy evolution as a function of the iterations. The inset presents the analytical ground state and optimized trial state. }
    \label{fig:opt-ho}
\end{figure}

A comparison between exact solutions and the results of the Monte Carlo simulation for the harmonic oscillator are displayed in \cref{fig:opt-ho}. Initially, the $M$ samples are drawn from a uniform distribution centred at the origin. Results obtained in this work are shown in blue dots and analytical results in dashed orange lines. The main graph shows the evolution of the energy as function of the optimization iterations, where the error bars are also included. The energy quickly reaches the GS energy as the simulation evolves. Moreover, the inset compares the trial wave function after the optimization procedure with the exact solution. For that, it was necessary to estimate the normalization factor $\mathcal N$ through a numerical quadrature integration. Notably, the optimized ANN shows excellent agreement with the exact solution.

\subsection{Morse oscillator}

This interatomic potential is often used to describe interactions between atoms, since the vibrational structure of molecules can be better approximated by the Morse oscillator than the harmonic oscillator. This is mainly due to the fact that this potential has unbound states and is anharmonic. Therefore, it also can describe bond breaking and account for the imbalance present in bonds. The potential is defined as

\begin{equation}
V(x) = D \left(
e^{-2ax} - 2 ~e^{-ax}
\right) ~,
\end{equation}

\noindent
where the natural length scale is $a_{\rm m}=\nf 1 a$, which implies that the energy units are given by $\nf {\hbar^2}{ma_{\rm m}^2}$. The well depth of the potential is considered to be $D=\nf 1 2$ for simplicity. In this units, the analytical GS energy is $E_0=-\nf 1 8$ and the dimensionless wave function is given by $\psi(\hat x)=\sqrt 2 \exp \left[- e^{-\hat x} - \nf {\hat x} 2\right]$.

\begin{figure}[tb]
    \centering
    \includegraphics[width=0.85\linewidth]{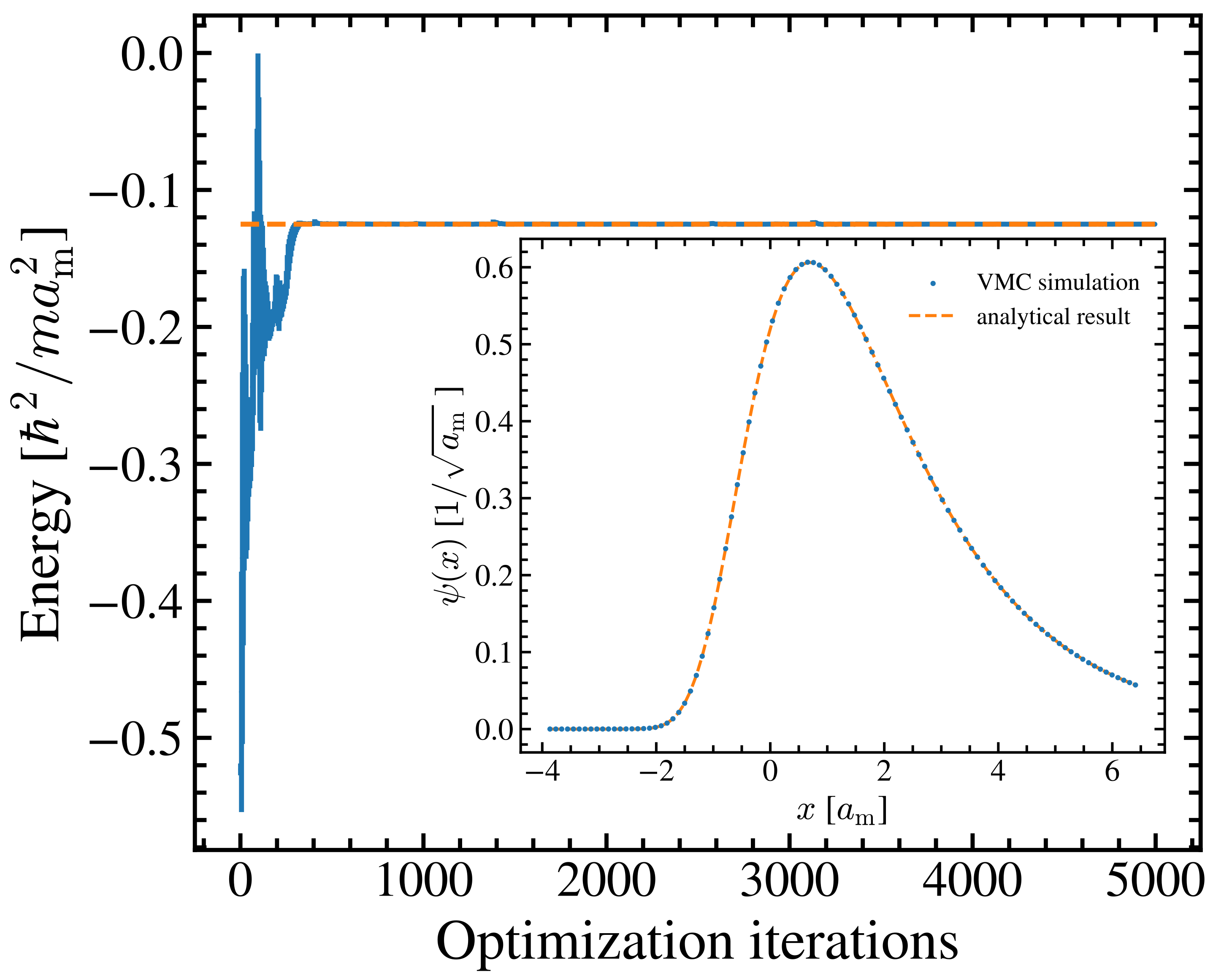}
    \caption{Minimization process for the Morse oscillator. The main graph shows the energy progression as a function of the number of iterations. The inset presents the analytical ground state and optimized trial state.}
    \label{fig:opt-mo}
\end{figure}

The results for ANN-based trial wave function description of the Morse oscillator are presented in \cref{fig:opt-mo}. The optimization procedure presents the system energy converging towards the GS energy $E_0=-0.125$, while the inset displays a comparison between the analytical and optimized trial wave functions, which shows excellent agreement. This result already hints on the versatility of ANNs as trial wave functions. With exactly the same parametrized class of functions $\phi_\theta$, accurate results were obtained for systems described by reasonably different potentials.

\subsection{Poschl-Teller potential}

To demonstrate this versatility even further, consider a class of potentials that are exactly solvable, smooth, and supports a tunable number of bound states given by the Poschl-Teller potential described by the form

\begin{equation}
V(x)=-~\frac {\hbar^2\lambda(\lambda + 1)}{2ma_{\rm pt}^2} {\rm sech}^2\left(\frac x {a_{\rm pt}}\right) ~,
\end{equation}

\noindent where $a_{\rm pt}$ is the natural length scale and, consequently, the energy units are given by $\nf{\hbar^2}{ma_{\rm pt}^2}$. The free parameter $\lambda$ controls the number of bound states. For the performed simulation it was considered only one bound state $\lambda=1$ and, therefore, the dimensionless GS energy is $E_0=-\nf 1 2$ and the analytical GS wave function is $\psi(\hat x)={\rm sech}(\hat x)$.

\begin{figure}[tb]
    \centering
    \includegraphics[width=0.85\linewidth]{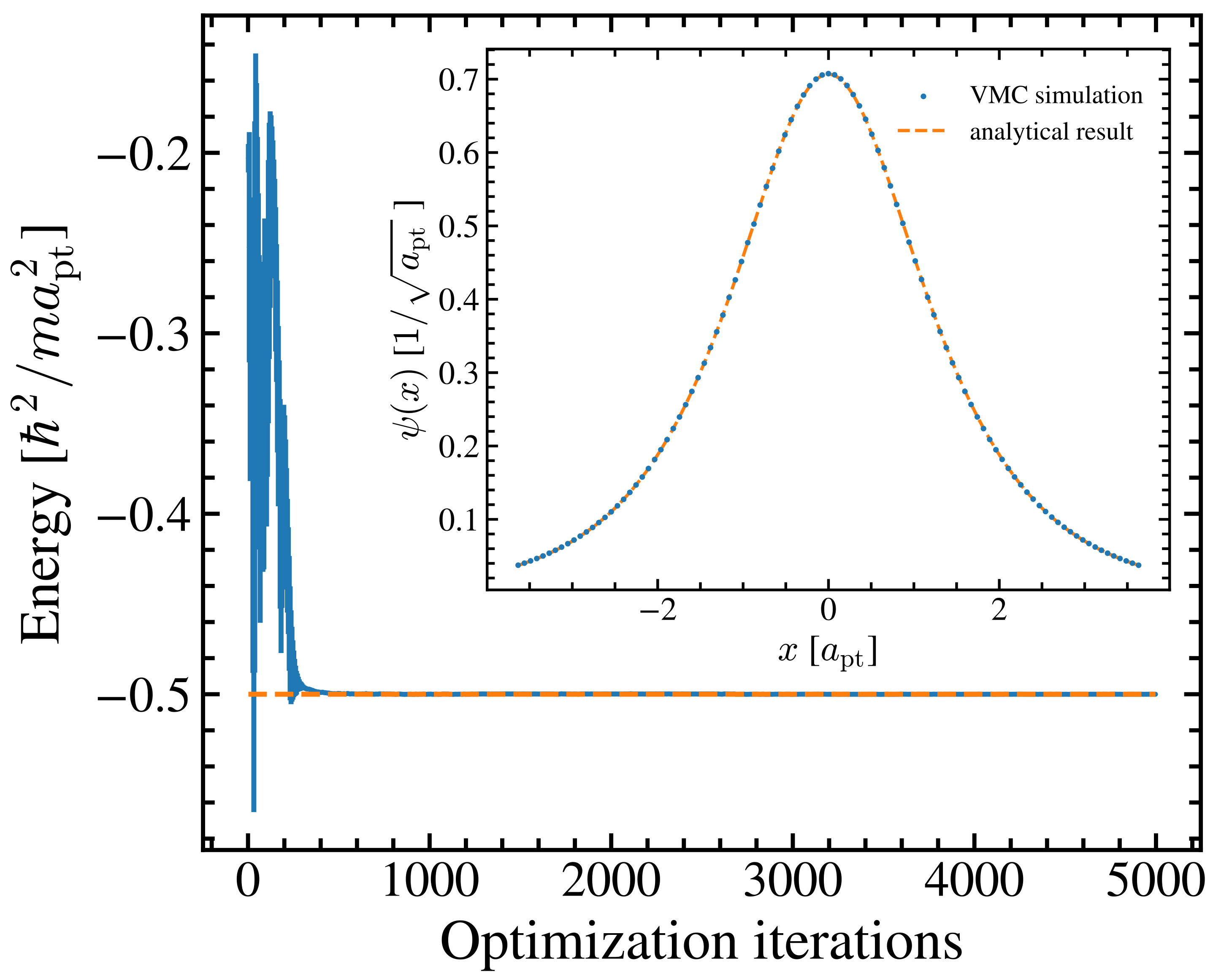}
    \caption{Minimization process for the Poschl-Teller potential. The main figure shows energy values as a function of optimization iterations. The inset presents the analytical ground state and optimized trial state.}
    \label{fig:opt-pt}
\end{figure}

This potential is often used in the context of atoms, molecules, and optics (AMO) physics, since it closely resembles realistic optical and trapping potentials, therefore, making it ideal for benchmarking. Once again, the evolution of the variational energy and its standard error as a function of the number of optimization steps is displayed in \cref{fig:opt-pt}, while the inset represents the comparison between analytical and optimized trial wave functions. The agreement among simulation and exact results is clear.

Hence, the ANN trial wave function was capable of accurately describing one-dimensional integrable systems. The next examples shall investigate systems with higher dimensions, particularly, all following examples will consider $D=3$. For single particle systems, that is $N=1$, the components are written as $\vct x\equiv\vct r=\{x,y,z\}$, while for two particle systems $\vct x\equiv\{\vct r_1,\vct r_2\}$, where $\vct r_i=\{x_i,y_i,z_i\}$.

\subsection{Hydrogen molecular ion}

As a more realistic example, consider the hydrogen molecular ion $H_2^+$, which is stabilized by a moving electron on the static field of two protons in their equilibrium positions, separated by $R=2.004~a_{\rm B}$. In chemistry, considering the protons of a molecule as static charges is known as the Born-Oppenheimer approximation, which approximates the kinetic energies of the protons to zero, since they are much heavier than electrons. Therefore, the potential energy is expressed as

\begin{equation}
V(\vct x) = - \frac {\nf {e^2}{4\pi\epsilon_0}}{\left|\vct r - \nf{\vct R}2\right|} -
\frac {\nf {e^2}{4\pi\epsilon_0}}{\left|\vct r + \nf{\vct R}2\right|} +
\frac {\nf {e^2}{4\pi\epsilon_0}}{R} ~,
\end{equation}

\noindent where $e$ is the fundamental charge, $\epsilon_0$ is the permittivity of vacuum, and $\vct R=R~\hat z$ is the vector that connects the two protons. For this and the final example, consider the system in atomic units, that is, Bohr radius as length scale and Hartree as energy units.

\begin{figure}[tb]
    \centering
    \includegraphics[width=0.85\linewidth]{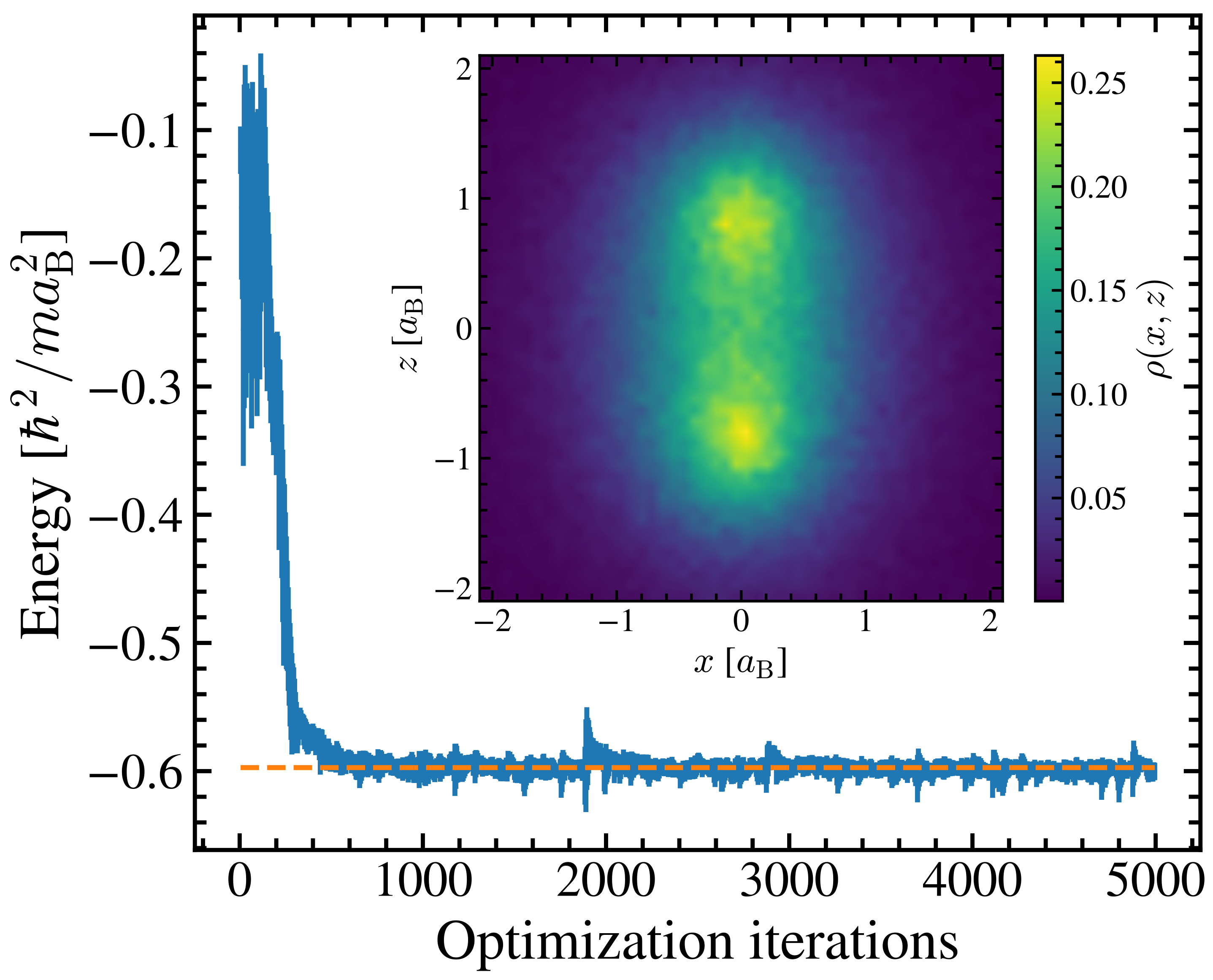}
    \caption{Description of H$_2^+$ by ANN-based trial functions. The minimization process is showing the evolution of the energy as a function of the number of iterations. The inset present a 2D projected distribution function. }
    \label{fig:opt-h2-ion}
\end{figure}

The solutions of this system are not analytically known, while numerical solutions predict that the GS energy is $-0.59724$ Hartree~\cite{sch96}. The simulation results for the evolution of the energy during the minimization process are displayed in~\cref{fig:opt-h2-ion} and compared with numerical results. Additionally, the inset shows the $xz$ projected PDF $\rho(x,z)=\int dy~|\psi_\theta(\vct r)|^2$, which demonstrate the larger bond due to the absence of one electron. Even without the information of the proton positions, the ANN was able to learn the GS function.

Therefore, quantitative results across all studied systems are summarized on~\cref{tab:ene}, where the energies are shown together with their respective statistical errors. The table also shows the reference energies that each system was compared to. 

\begin{table}[tb]
\caption{Comparison between the optimized (columns two and five) and reference (columns three and six) energy values across the studied systems (columns one and four).}
\begin{tabular}{ccccccc} \hline\hline
        System          & ANN-VMC     && Ref.     & System          & ANN-VMC     & Ref.     \\ \hline\hline
        $E_{\rm HO}$    & 0.5000(1)   && 0.5      & $E_{\rm Yu}$    & -0.322(5)   & -0.326   \\
        $E_{\rm MO}$    & -0.12500(2) && -0.125   & $E_{\rm H^+_2}$ & -0.595(5)   & -0.59724 \\
        $E_{\rm PT}$    & -0.50001(5) && -0.5     & $E_{\rm H_2}$   & -1.16(1)    & -1.1645  \\
        \hline\hline
\end{tabular}
\label{tab:ene}
\end{table}



\end{document}